\documentclass[10pt]{article}
\usepackage{ucs} 
\usepackage[utf8x]{inputenc}
\usepackage{amsmath,amsfonts,amssymb,amsthm,amsopn,amscd}
\usepackage{bbm}
\usepackage{graphicx}
\usepackage[colorlinks=true]{hyperref}

\usepackage[labelfont=bf,labelsep=period,justification=raggedright]{caption}

\makeatletter
\renewcommand{\@biblabel}[1]{\quad#1.}
\makeatother

\date{}

\theoremstyle{plain}
\newtheorem*{theorem*}{Theorem}

\theoremstyle{definition}
\newtheorem*{proposition*}{Proposition}
\newtheorem*{definition*}{Definition}

\newcommand{\bd}{\begin{document}}
\newcommand{\ed}{\end{document}}
\newcommand{\beq}{\begin{equation}}
\newcommand{\eeq}{\end{equation}}
\newcommand{\nid}{\noindent}
\newcommand{\ben}{\begin{enumerate}}
\newcommand{\een}{\end{enumerate}}
\newcommand{\bit}{\begin{itemize}}
\newcommand{\eit}{\end{itemize}}
\newcommand{\baR}{\begin{array}}
\newcommand{\eaR}{\end{array}}
\newcommand{\su}{\section}\usepackage{graphicx}
\newcommand{\ssu}{\subsection}
\newcommand{\sssu}{\subsubsection}
\newcommand{\deq}{\stackrel {\rm def}{=}}
\newcommand{\bea}{\begin{eqnarray}}
\newcommand{\eea}{\end{eqnarray}}

\newcommand{\R}{\mathbb{R}}  
\newcommand{\Z}{\mathbbm{Z}}  
\newcommand{\C}{\mathbbm{C}}  
\newcommand{\mr}{\mathbf{r}}  
\newcommand{\me}{\mathbf{e}}
\newcommand{\mk}{\mathbf{k}}
\newcommand{\T}{\mathcal{T}} 
\newcommand{\mH}{\mathcal{H}}  
\newcommand{\gt}{\tau} 
\newcommand{\F}{{\mathcal F}}
\newcommand {\image} {\includegraphics}

\newenvironment{heuriproof}{\noindent \emph{Heuristic proof.}}{\begin{flushright} \qed \end{flushright}}
\newcommand{\eqdef}{\overset{\rm def}{=}}

\graphicspath{{figures/}}

\begin{document}
\begin{flushleft}
{\Large
\textbf{Hyperbolic planforms in relation to visual edges and textures perception}
}
\\
Pascal Chossat$^{1}$, 
Olivier Faugeras$^{2,\ast}$
\\
\bf{1} Pascal Chossat Dept. of Mathematics, University of Nice Sophia-Antipolis, JAD Laboratory and CNRS, Parc Valrose, 06108 Nice Cedex 02, France\\
	Also NeuroMathComp Laboratory
\\
\bf{2} Olivier Faugeras NeuroMathComp Laboratory, INRIA/ENS Paris/CNRS, 2004 Route des Lucioles, 06902 Sophia-Antipolis, France
\\
$\ast$ E-mail: Olivier.Faugeras@sophia.inria.fr
\end{flushleft}

\section*{Abstract}
We propose to use bifurcation theory and pattern formation as theoretical probes for various hypotheses about the neural organization of the brain. This allows us to make  predictions about the kinds of patterns that should be observed in the activity of real brains through, e.g. optical imaging, and opens the door to the design of experiments to test these hypotheses. We study the specific problem of visual edges and textures perception and suggest that these features may be represented at the population level in the visual cortex as a specific second-order tensor, the structure tensor, perhaps within a hypercolumn. We then extend the classical ring model to this case and show that its natural framework is the non-Euclidean hyperbolic geometry. This brings in the beautiful structure of its group of isometries and certain of its subgroups which have a direct interpretation in terms of the organization of the neural populations that are assumed to encode the structure tensor. By studying the bifurcations of the solutions of the structure tensor equations, the analog of the classical Wilson and Cowan equations, under the assumption of invariance with respect to the action of these subgroups, we {\em predict} the appearance of characteristic patterns. These patterns can be described by what we call hyperbolic or H-planforms that are reminiscent of Euclidean planar waves and of the planforms that were used in \cite{bressloff-cowan-etal:01,bressloff-cowan-etal:02} to account for some visual hallucinations. If these patterns could be observed through brain imaging techniques they would reveal the built-in or acquired invariance of the neural organization to the action of the corresponding subgroups.\vspace{-0.2cm}

\section*{Author summary}
Naive introspection conveys to us the vivid feeling that our visual perception of the outside world is remarkably stable and invariant despite the fact that we move our gaze and body. This must be the effect of the neuronal organization of the visual areas of our brains that, despite the high variability of the flux of photons impinging on our retinas, manage to maintain in our consciouness a representation that seems to be protected from brutal variations. In this article we propose a theory to account for an invariance that pertains to such image features as edges and textures. The theory is based on the simple assumption that the spatial variations of the image intensity, also called its derivatives, are extracted and represented in such brain areas as the hypercolumns of V1 by populations of neurons that excite and inhibit each other according to the values of these derivatives. Geometric transformations of the retinal image, caused say by eye movements, affect these derivatives and their neuronal representations. Assuming that these representations are invariant to these transformations we predict the appearance of some specific patterns of activity which can be described by what we call hyperbolic planforms. These hyperbolic planforms correspond to the usual planar waves or planforms that have been used in \cite{bressloff-cowan-etal:01,bressloff-cowan-etal:02} to account for some visual hallucinations, and arise naturally from our assumptions about the way the image derivatives are represented in neural populations and about their invariance to some retinal transformations. It is a surprising feature of our work that the natural geometry that emerges from our work is not the usual Euclidean geometry we are all used to but the much less familiar hyperbolic, non-Euclidean, geometry that was made famous by the work of Lobatchewsky. We also propose some preliminary ideas for  putting our theory to test by actual measurements of brain activity.
\section*{Introduction}
Visual perception, computational or biological, depends upon the extraction from the raw flow of images incoming on the retina of a number of image features such as edges, corners, textures or directions of motion, at a variety of spatio-temporal scales. All these features involve comparing some functions of the incoming intensity values at nearby spatio-temporal locations and this points very strongly to the notion of derivatives. The idea of constructing the image representations from various derivatives of the intensity flow is at the heart of the concept of the primal sketch put forward in the seventies by the late David Marr \cite{marr:82} or the concept of $k$-jets borrowed from mathematics by Jan Koenderink and his colleagues \cite{koenderink-doorn:87,florack-romeny-etal:96}. A quick look at the computer vision or image processing literatures will convince anyone of the universal use of image derivatives in feature extraction algorithms \cite{pratt:78,ballard-brown:82,horn:86,forsyth-ponce:03} . There is also strong evidence that the visual system of many species is organized in such a way that quantities related to image derivatives
are extracted, and hence represented, by neuronal activity \cite{chalupa-werner:04}. The notion of derivative is misleading though because it often implies in people's minds the idea of linearity. But of course it does not have to be the case, computer vision algorithms are usually highly nonlinear even if they use derivatives, and nonlinearities are omnipresent in the brain and in the parts of it that are dedicated to visual perception.

If we accept these two ideas, 1) that image derivatives are represented in the visual pathway and 2) in a nonlinear fashion, this immediately raises the related questions of the coordinate system(s) in which they are represented and the effect of changing such coordinate system(s). Changes of coordinate systems are described by group actions such as those of the familiar groups of translations and rotations in the Euclidean plane. This leads  naturally to the idea of group invariance: one can argue that the image features representations should be somewhat robust to these groups actions. This is of course only a hypothesis albeit a likely one, we think. In computer vision this idea is not new and there was a time when a significant part of this community was actively designing feature representations that were invariant with respect to a variety of group actions \cite{azores:93}. What is interesting in the case of biological vision is that this hypothesis has consequences that may be testable experimentally: If the visual pathway is organized so as to support invariance of feature representations at the mesoscopic level, say the hypercolumn in V1, we may be able to predict the appearance of certain patterns of activity in the involved neuronal populations that are a direct consequence of the invariance hypothesis.

In this article we begin the development of a mathematical theory of the processing of image edges and textures in the hypercolumns of area V1 that is based on a nonlinear representation of the image first order derivatives called the structure tensor. Assuming that this tensor is represented by neuronal populations in the hypercolumns of V1 that interact in a way that can be described by equations similar to those proposed by Wilson and Cowan \cite{wilson-cowan:73}, bifurcation theory allows us to predict  the formation of specific patterns in the cortical medium that are related to the assumed invariant properties of the underlying cortical representation of the structure tensor.

%
%
\section*{Methods}
\subsection*{The structure tensor as a representation of edges and textures}\label{section:structuretensor}
The structure tensor is a way of representing the edges and texture of a 2D image $I(x,y)$ \cite{bigun-granlund:87,knutsson:89}. 
Let $g_{\sigma_1}(x,y)=\frac{1}{2 \pi \sigma_1^2} exp(-(x^2+y^2)/2\sigma_1^2)$ be the two-dimensional Gaussian function with 0 mean and variance $\sigma_1^2$. We consider the regularized image $I_{\sigma_1}$ obtained by convolving the image $I$ with $g_{\sigma_1}$, we note $I_{\sigma_1}=g_{\sigma_1} \star I$, where the symbol $\star$  represents the convolution operation. The gradient $\nabla I_{\sigma_1}$ of $I_{\sigma_1}$ is a two-dimensional vector which emphasizes image edges: within a flat region $\nabla I_{\sigma_1}=0$, at a pronounced edge $\| \nabla I_{\sigma_1} \|$, the Euclidean norm of $\nabla I_{\sigma_1}$ is large, and $\nabla I_{\sigma_1}$ points in the normal direction of the edge. The parameter $\sigma_1$ is called the local scale. One then forms the $2 \times 2$ symmetric matrix $\T_0(\nabla I_{\sigma_1})=\nabla I_{\sigma_1} \otimes \nabla I_{\sigma_1}=\nabla I_{\sigma_1} \,^t\hspace{-0.025cm}\nabla I_{\sigma_1}$, where $\otimes$ indicates the tensor product and $\ ^t$ indicates the transpose of a vector. By convolving $\T_0(\nabla I_{\sigma_1})$ componentwise with a Gaussian $g_{\sigma_2}$ we obtain the matrix $\T_{\sigma_2}(\nabla I_{\sigma_1})=g_{\sigma_2} \star \T_0(\nabla I_{\sigma_1})$. It is not hard to verify that this symmetric matrix is positive, i.e. $^t\hspace{-0.05cm}z\hspace{.05cm} \T_{\sigma_2}(\nabla I_{\sigma_1}) z \geq 0$ for all vectors $z$ in $\R^2$. It is called the structure tensor. When there is no ambiguity we will use $\T$ instead of $\T_{\sigma_2}(\nabla I_{\sigma_1})$.

Note that the construction of the structure tensor involves two spatial scales. The first one, defined by $\sigma_1$, is the one at which the image derivatives are estimated. The structure tensor is insensitive to noise and irrelevant details at scales smaller than $\sigma_1$. The second one, defined by $\sigma_2$, is the one at which the averages of the estimates of the image derivatives are computed, it is the integration scale, and is related to the characteristic size of the texture to be represented, and to the size of the receptive fields of the neurons that may represent the structure tensor.

Being symmetric and positive, $\T$ has two orthonormal eigenvectors $\me_1$ and $\me_2$ and two positive corresponding eigenvalues $\lambda_1$ and $\lambda_2$ which we can always assume to be such that $\lambda_1 \geq \lambda_2 \geq 0$. The distribution of these eigenvalues in the $(\lambda_1,\lambda_2)$ plane reflects the local organization of the image intensity variations. Indeed, one can establish a correspondence between local intensity patterns and relative values of $\lambda_1$ and $\lambda_2$. For example constant areas are characterized by $\lambda_1 =\lambda_2 =0$, straight edges give $\lambda_1 >\hspace{-0.1cm}> \lambda_2 \simeq 0$, their orientation being that of $\me_2$, corners yield $\lambda_1 \geq \lambda_2  >\hspace{-0.1cm}> 0$. The difference $\lambda_1-\lambda_2$ becomes large for anisotropic textures. These simple examples are intended to show the richness of the structure tensor when it comes to representing textures and edges at a given spatial scale.

This representation of the local image orientations and textures is richer than, and contains, the local image orientations model which is conceptually equivalent to the direction of the local image intensity gradient $g_{\sigma_2} \star \nabla I_{\sigma_1}$. The local image orientation is a one-dimensional representation which can be obtained from the local image intensity gradient, which is two-dimensional, as the ratio of the gradient components. The structure tensor itself is three-dimensional. Its three dimensions can be either pictured as its three entries or as the collection of its two eigenvalues and the direction of one of its eigenvectors, e.g. the one corresponding to the largest eigenvalue. In particular, it should be clear from the above that the structure tensor can discriminate local intensity patterns that would be otherwise confused by the local orientations model: For example, given an isotropic structure localized in an image neighbourhood of size of the order of the integration scale $\sigma_2$  with  no preferred direction of gradient, the local gradients average out resulting in a zero magnitude. An example of such an isotropic structure is a black disk of diameter $\sigma_2 /2$ on a white background. There is clearly gradient information; however, since there is no preferred phase, it zeros itself out as in the case of a uniformly grey pattern. The eigenvalues of the structure tensor turn out to be both equal to some strictly positive number in the case of the disk and both equal to 0 in the case of the uniformly grey pattern. This is an extreme example but one may also think of a texture pattern made of short line elements pointing in roughly the same direction. The local gradients average to a direction roughly perpendicular to the average direction of the line elements. The length of the resulting vector is an indication of the average contrats across these line elements. In the case of the structure tensor, the unit eigenvector, together with its corresponding largest eigenvalue, contains the same information but the second eigenvalue contains information about the spread in the directions of the line elements, the difference between the two eigenvalues being, as mentioned above, an indication of the anisotropy of the texture. This discussion should have made it clear that the structure tensor contains, at a given scale, more information than the local image intensity gradient at the same scale.

The question of whether some populations of neurons in such a visual area as V1, can represent the structure tensor cannot be answered at this point in a definite manner but we hope that the predictions of the theory we are about to develop will help deciding on this issue. We can nonetheless argue as follows. We know that orientation hypercolumns in V1 represent local edge orientations in receptive fields whose size vary between 0.5 and 2 degrees. This corresponds to values of $\sigma_2$ between 0.5 and 2 centimeters at a viewing distance of 57 centimeters. For a given orientation $\theta$, the two orientations $\theta+\pi/4$ and $\theta+\pi/2$ are also represented in the orientation hypercolumn and this is very much the same as representing the three components of the stucture tensor at this scale. Indeed, let us denote by $\nabla I^{\theta}_{\sigma_1}$ the component of the smooth gradient in the directions $\theta$. It is easy to show that $\nabla I^{\theta+\pi/4}_{\sigma_1}=\frac{1}{\sqrt{2}}\left(\nabla I^{\theta}_{\sigma_1}+\nabla I^{\theta+\pi/2}_{\sigma_1}\right)$ and it follows that the product $\nabla I^{\theta}_{\sigma_1}\nabla I^{\theta+\pi/2}_{\sigma_1}$ is a linear combination of $\left(\nabla I^{\theta}_{\sigma_1}\right)^2$, $\left(\nabla I^{\theta+\pi/4}_{\sigma_1}\right)^2$, and $\left(\nabla I^{\theta+\pi/2}_{\sigma_1}\right)^2$. This remains true of the local averages of these quantities obtained by convolution with the Gaussian of standard deviation $\sigma_2$. We note that these three components are represented in the Euclidean coordinate system defined by the orientation $\theta$ and the orthogonal direction $\theta+\pi/2$. So we may say that the joint activity of the populations of neurons in the hypercolumn representing these three orientations is in effect an encoding of the structure tensor. This reasoning applies to any orientation $\theta$ and it follows that the joint activity of all triplets of populations of neurons in the hypercolumn that encode the triplets of orientations $(\theta, \theta+\pi/2, \theta+\pi/4)$ for all possible values of $\theta$ between 0 and $\pi$ are a representation of the structure tensor that is roughly invariant to the choice of the orientation of the coordinate system in which it is represented or more accurately that contains all such representations which differ by a rotation of the coordinate system, up to the accuracy of the orientation representation in the orientation hypercolumn.
Where in V1 could one find populations of neurons that encode the structure tensor? Cytochrome oxydase (CO) blobs and their neighbourhoods seem to be good candidates since their distribution appears to be correlated with a number of periodically repeating feature maps in which local populations of neurons respond preferentially to stimuli with particular properties such as orientation, spatial frequency, brightness and contrast \cite{blasdel:92,blasdel-salama:86,bonhoeffer-grinvald:91,bonhoeffer-kim-etal:95,issa-trepel-etal:00,kaplan:04,casagrande-xu:04}. It has thus been suggested that the CO blobs could be the sites of functionally and anatomically distinct channels of visual processing \cite{edwards-purpura-etal:95,livingstone-hubel:84,sincich-horton:02,tootell-hamilton-etal:88}. Recently Bressloff and Cowan \cite{bressloff-cowan:03,bressloff-cowan:03b} introduced a model of a hypercolumn in V1 consisting of orientation and spatial frequency preferences organized around a pair of pinwheels. One pinwheel is centered at a CO blob and encodes coarse to medium coarse scales, the other is centered at a region that encodes medium coarse to fine scales. Despite the fact that these authors do not consider the encoding of brightness and contrast, it has been suggested by other authors \cite{allman-zucker:90} that this might also be the case. Such a hypercolumn is therefore a good candidate for representing the structure tensor at several scales as well as, as these authors claim, the local orientations at various spatial frequencies.
As a consequence of this discussion we assume that the structure tensor is represented by the activity of the populations of neurons in a hypercolumn, where the word represented is to be understood as explained above.


Let therefore $\T$ be a structure tensor. We assume that there is some quantity which we associate to an average membrane potential, noted $V(\T,\tau)$, and is a function of $\T$ and the time $\tau$ abd which is, e.g., high if $\T$ reflects the actual intensity values in the column receptive fields and low otherwise. We assume that its time evolution is governed by an equation of the Wilson and Cowan \cite{wilson-cowan:73} or Amari \cite{amari:77} type.
\begin{equation}\label{eq:neuralmass}
 V_\tau(\T,\tau)=-\alpha V(\T,\tau)+\int_\mH w(\T,\T')S(V(\T',\tau))\,d\T'+I(\T,\tau),
\end{equation}
where the integral is taken over $\mH$, the set of possible structure tensor. We provide below a precise mathematical definition of this set. $d\T'$ is the corresponding area element, also defined below, and $I$ is an input current. 

The positive coefficient $\alpha$ can be normalized to 1 by a suitable choice of time scale. $S$ is a sigmoidal function which after normalization may be expressed as:
\begin{equation}\label{eq:sig}
S(x)=\frac{1}{1+e^{-\mu x}} \quad x \in \R,
\end{equation}
where $\mu$ is a positive coefficient which governs the stiffness of the sigmoid.

The function $w$. called the connectivity function, is defined as follows.
If we assume further that the neuronal population representing the value $\T$ of the structure tensor excites (respectively inhibits) the neuronal population representing the value $\T'$ if the distance $d(\T,\T')$  is small (respectively large), a natural form of the connectivity function $w$ is obtained from the following function $g$, a difference between two pseudo-Gaussians:
\begin{equation} \label{fonction:g}
 g(x)=\frac{1}{\sqrt{2\pi \sigma_1^2}}  e^{\displaystyle -\frac{f(x)}{2\sigma_1^2}}-\theta \frac{1}{\sqrt{2\pi \sigma_2^2}} e^{\displaystyle -\frac{f(x)}{2\sigma_2^2}},
\end{equation}
where $\sigma_1 < \sigma_2$, $\theta \leq 1$, and $f$ is a monotonously increasing function from the set $\R^+$ of positive real numbers to $\R^+$. For example, if $f(x)=x^2$ we obtain the usual difference of Gaussians. 

One then defines
\[
 w(\T,\T')=g(f(d(\T,\T')))
\]
$w$ is clearly invariant to the action of the isometries $\gamma$ of $\mH$:
\[
 w(\gamma \cdot \T, \gamma \cdot \T')=w(\T,\T') \quad \forall\, \gamma 
\]
We will see that with such a choice of connectivity function, the integral in \eqref{eq:neuralmass} is well-defined because $w$ is small at ``infinity''.

This is similar in spirit to the ring model described in \cite{hansel-sompolinsky:97,ermentrout:98}, see the Discussion Section.

There are of course many loosely defined terms in the presentation so far, including the definition of the set of structure tensors, of the distance between two such tensors that plays a central role in the construction of the connectivity function $w$, and the definition of the isometries of the set of structure tensors, i.e. the transformations that leave the distance between two tensors unchanged. We provide below precise answers to all these questions. Before doing this we explain how equation \eqref{eq:neuralmass} which describes the dynamics of a neural mass, e.g. a hypercolumn of V1, can be ``spatialized'' in order to provide
a neural or cortical field model (see \cite{deco-jirsa-etal:08,ermentrout:98} for reviews of neural fields) that could describe the spatio-temporal activity of V1 related to the representation of edges and textures.

Indeed let us assume the existence a continuous distribution of such columnar systems in a regular bounded open set $\Omega$ of $\R^2$, modeling a piece of a flat cortex. We note $\mr$ the spatial variable. Equation \eqref{eq:neuralmass} can be generalized to the following
\begin{equation}\label{eq:neuralfield}
 V_\tau(\mr,\T,\tau)=-V(\mr,\T,\tau)+\int_\Omega\int_\mH w(\mr,\T,\mr',\T') S(V(\mr',\T',\tau))\,d\T'\,d\mr'+I(\mr,\T,\tau),
\end{equation}
where $d\mr'$ is the usual Euclidean area element. The average membrane potential $V$ depends on the position $\mr$ in the continuum, i.e. on the position of the hypercolumn in V1, on the time $\tau$ and on the possible local values of the structure tensor $\T$. The connectivity function $w$ is now a function of the structure tensors $\T$ at point $\mr$ of the continuum and $\T'$ at point $\mr'$.

We do not deal any further with this equation, leaving it for future work.

Considering equation \eqref{eq:neuralmass} we will study how its solutions change when the slope parameter $\mu$ increases from the value 0. This study, together with the formulation of hypotheses about the invariance
of the average membrane potential with respect to the action of some subgroups of the group of isometries of the set of structure tensors, predicts, through bifurcations of the solutions to \eqref{eq:neuralmass}, the appearance of certain patterns displaying the kind of symmetries described by these subgroups. If such patterns can indeed be observed by actual measurements, e.g., optical imaging \cite{grinvald-hildesheim:04}, then this would be a strong indication that the neural ``hardware'' is built in such a way that its state is insensitive to the action of these subgroups. To say things differently, bifurcation theory and pattern formation could potentially become theoretical probes for the validity of various hypotheses about the neural organization of the brain, allowing to make predictions about the kinds of patterns that should be observed in the activity of real brains, and opening the door to the design of experiments to test these hypotheses. This is indeed an exciting perspective. We now proceed to flesh  up the theory.
\subsection*{The mathematical structure of the set of structure tensors}
We present some important properties of the set of structure tensors. These properties are somewhat scattered in the literature and are relevant to our forthcoming discussion of pattern formation in cortical tissues. 

The key observation is that the structure tensors naturally live in a hyperbolic space of dimension 3 that can be peeled, like an onion, into sheets of dimension 2, each sheet corresponding to a constant value of the determinant of the elements inhabiting it. We are therefore led to study  hyperbolic spaces of dimension 2 which turn out to enjoy a very simple representation in the open unit disk $D$ of the complex plane, the so-called Poincaré disk, with its fascinating non-Euclidean geometry that arises from the Riemannian structure of the set of structure tensors. This geometry has been studied in depth by mathematicians and theoretical physicists and is still a very active research area with many open difficult questions. We then establish the dictionary that will allow us to translate statements about the structure tensors of determinant equal to one into statements about complex numbers of magnitude less than or equal to 1. The fundamental new item in this section is the group of isometries of the Poincaré disk, analog to the group of rigid displacements in the Euclidean plane, whose action on complex numbers can be translated (the technical word is lifted) into meaningful actions on structure tensors. We explain in the supplementary text S1 how to put things back together, that is to say, how to reconstruct in a mathematically coherent fashion the onion representing the whole set of structure tensors from the description of one of its sheets, or peels, i.e. the one corresponding to the unit determinant structure tensors. The final  touch is a somehow deeper analysis of some subgroups of the group of isometries of $D$ introduced previously. These subgroups arise naturally when one examines the kinds of invariances that the cortical representations of the structure tensors should enjoy. The mathematical structure that emerges in this context is that of a Fuchsian group, introduced by Henri Poincaré in 1882 \cite{poincare:82}.

Consider the set ${\rm SDP}(2)$ of $2 \times 2$ symmetric positive-definite matrices (see glossary in table \ref{table:glossary}). 
Indeed, let
\begin{equation}\label{eq:tensorabc}
 \T=    \left[
\begin{array}{cc}
 a & c\\
c & b
\end{array}
\right],\,a>0,\,ab-c^2 > 0
\end{equation}
be an element of ${\rm SDP}(2)$. 

We refer to $a$ (respectively $b$, $c$) as the $a$-coordinate (respectively the $b$- $c$-coordinate) of $\T$.

If we scale $\T$ by $\lambda > 0$, $\lambda \T$ is also an element of ${\rm SDP}(2)$. Hence ${\rm SDP}(2)$ is a positive cone. It is open because it is defined by two strict inequalities.

It is also a three-dimensional Riemannian manifold in which the distance is defined as follows \cite{moakher:05}. \\
Given $\mathcal{T}_1$ and $\mathcal{T}_2$ in ${\rm SDP}(2)$, the Riemannian distance $d_0(\mathcal{T}_1,\mathcal{T}_2)$ can be expressed as the Frobenius norm\footnote{The Frobenius norm of a real matrix is the square root of the sum of the squares of its elements.} of the  principal logarithm of $\mathcal{T}_1^{-1} \mathcal{T}_2$:
\begin{equation}\label{eq:d0}
d_0(\mathcal{T}_1,\mathcal{T}_2)=\| \log \T_1^{-1} \T_2 \|_F=\left( \sum_{i=1,2}  \log^2 \lambda_i \right)^{1/2},
\end{equation}
where the $\lambda_i$s are the eigenvalues of the matrix $\T_1^{-1} \T_2$.
This expression is symmetric with respect to $\mathcal{T}_1$ and $\mathcal{T}_2$ since $\mathcal{T}_2^{-1}\mathcal{T}_1=\left(\mathcal{T}_1^{-1} \mathcal{T}_2\right)^{-1}$ and the $\lambda_i$s are positive since $\mathcal{T}_1^{-1} \mathcal{T}_2$ is conjugate to the symmetric positive definite matrix $\mathcal{T}_2^{1/2} \mathcal{T}_1^{-1} \mathcal{T}_2^{1/2}$.

This definition of the distance between two tensors can be motivated  from a biological viewpoint. A tensor is a symmetric $2 \times 2$ matrix, hence it can be thought of a a three-dimensional vector $(a,b,c)$. The ``natural'' distance between two such vectors (representing the tensors $\T_1$ and $\T_2$) is the usual Euclidean distance $(a_1-a_2)^2+(b_1-b_2)^2+(c_1-c_2)^2$. This distance has the following problem. A tensor $\T$ defines a quadratic form $z \to ^t\hspace{-0.05cm}z \T z$. If we change the coordinate system in which we express the coordinates of two tensors $\T_1$ and $\T_2$  they become $^t\hspace{-0.03cm}M \T_1 M$ and $^t\hspace{-0.003cm}M \T_2 M$, where $M \in {\rm GL}(2,\R)$ is the matrix defining the change of coordinate system. It can be verified that this transformation does not leave in general the Euclidean distance invariant whereas it does leave $d_0$ invariant. This invariance is a very desirable feature since the measure of similarity between two tensors (their distance) should not depend on the particular coordinate system used to evaluate their components. Hence it is very likely that evolution would rather select $d_0$ than the simpler but sometimes misleading Euclidean distance.

\begin{figure}[!ht]
 \centerline{
\includegraphics[width=4in]{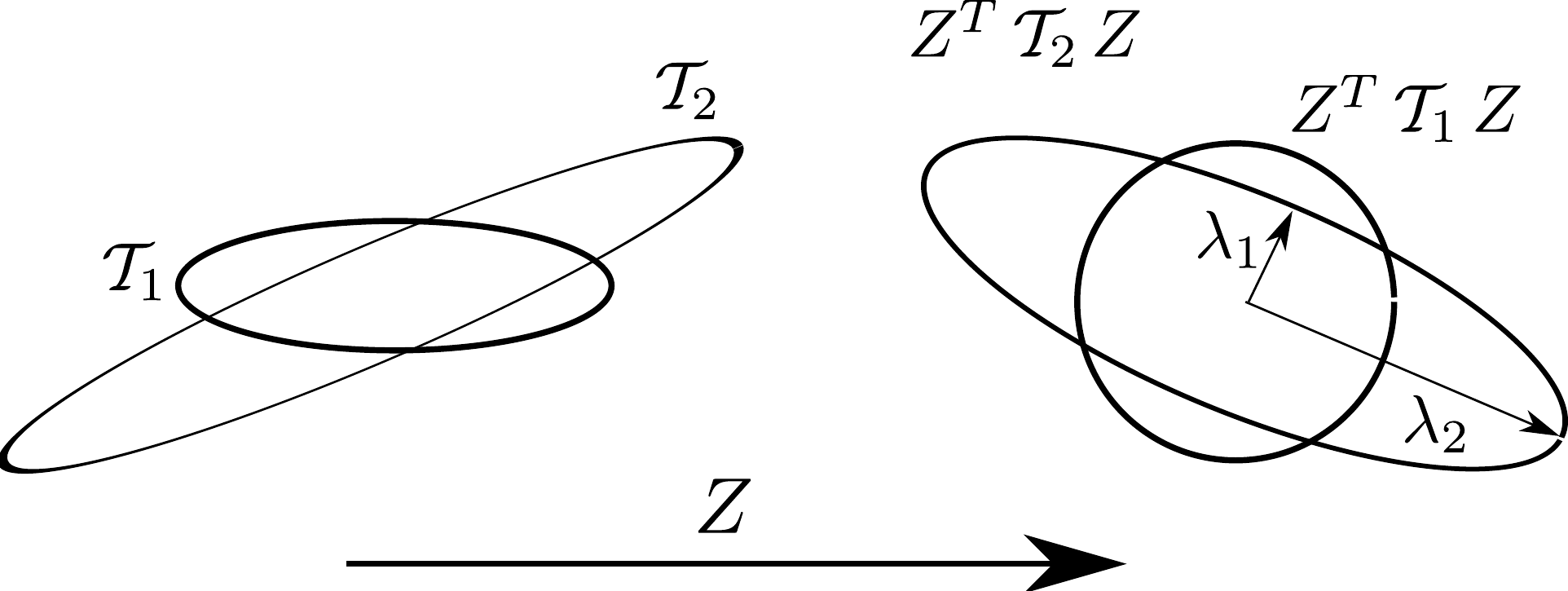}
}
\caption{{\bf The two structure tensors $\T_1$ and $\T_2$ are represented by the elliptic blobs shown in the lefthand side of the figure. After the change of coordinates defined by the matrix $Z$, $\T_1$ is represented by the unit disk and the principal axes of $\T_2$ are equal to the eigenvalues $\lambda_1$ and $\lambda_2$ that appear in \eqref{eq:d0}, see text.}}
\label{fig:tensor_distance}
\end{figure}

From yet another perspective it can be shown, see e.g. \cite[Volume 1, Chapter X, Theorem 9]{gantmacher:59}, that there exists a change of coordinates, i.e., a $2 \times 2$ matrix $Z$ such that in the new coordinate system $^t\hspace{-0.05cm}Z \T_2 Z={\rm diag}(\lambda_1,\lambda_2)$ and $^t\hspace{-0.05cm}Z\T_1Z={\rm Id}_2$. In other words, the distance \eqref{eq:d0}, is a measure of how well $\T_1$ and $\T_2$ can be {\em simultaneously} reduced to the identity matrix by a change of coordinate system. This change of coordinate system is {\em not} in general a pure rotation but a combination of a pure rotation and a scaling of the coordinates. If we picture the structure tensor $\T$ as the elliptic blob defined by the equation $^t\hspace{-0.05cm}z \T z \leq 1$, $z=(x,y)$, the two tensors $\T_1$ and $\T_2$ are represented by two elliptic blobs as shown in the lefthand part of figure \ref{fig:tensor_distance}. After the coordinate transform defined by $Z$, $\T_1$ is represented by a unit disk and $\T_2$ by an elliptic blob whose major axes are the eigenvalues $\lambda_1$ and $\lambda_2$ that appear in \eqref{eq:d0}, as shown in the righthand part of the same figure.
There is a unique geodesics (curve of shortest length) between two elements of ${\rm SDP}(2)$. Its expression is given in the supplementary text material S3.

If we now consider the two-dimensional submanifold ${\rm SSDP}(2)$ of the special positive definite matrixes whose determinant $ab-c^2$ is equal to 1, it is clear that ${\rm SDP}(2)={\rm SSDP}(2) \times \R^+$.  We detail this point in the supplementary text material S1. 

It can be shown that ${\rm SSDP}(2)$ equiped with the Riemannian metric induced by that of ${\rm SDP}(2)$ is a Riemannian surface with constant sectional curvature equal to -1, see the supplementary text material S1 for details. This indicates that it is isomorphic to the two-dimensional hyperbolic space, noted $H^2$, for which we now provide three different models.

There are three main models of $H^2$, the two-dimensional hyperbolic space. Each model has its advantages and disadvantages. We first present the hyperboloid model which is
the most natural for the set of structure tensors, next the Poincaré disk model which is the most convenient for carrying out analytic computations. We relegate in the supplementary text material S2 the third model, called the Poincaré half-plane model and noted $\mathcal H$, which is not
as convenient as the second for visualizing important geometric transformations such as rotations.

The hyperboloid model is defined as the hyperboloid sheet in $\R^3$ of equation
\[
 x_0^2-x_1^2-x_2^2=1, \quad x_0 >0,
\]
associated to the quadratic form $q(x)=x_0^2-x_1^2-x_2^2$ which yields by polarization the bilinear form $b(x,x')=x_0 x_0'-x_1 x_1'-x_2 x_2'$. The corresponding Riemannian distance
is given by
\[
 d_1(x,x')={\rm arccosh}\, b(x,x').
\]
Geodesics are the curves intersections of the hyperboloid sheet with planes through the origin.

The Poincaré disk model is conveniently obtained by stereographic projection on the plane of equation $x_0=0$ through the point of coordinates $(-1,0,0)$ of the hyperboloid model. This establishes a one to one mapping of the hyperboloid sheet onto the open unit disk $D$. Given two points $z$ and $z'$ of $D$ corresponding to the points $x$ and $x'$ of the hyperboloid, the corresponding Riemannian distance is given by
\newcommand{\arctanh}{{\rm arctanh}}
\beq \label{d2}
 d_2(z,z')=\arctanh \frac{|z-z'|}{|1-\overline zz'|},
\eeq
and satisfies $d_2(z,z')=d_1(x,x')$. 
We may also write
\beq \label{d2-bis}
d_2(z,z') = \frac{1}{2}\log \frac{|1-\overline zz'|+|z'-z|}{|1-\overline zz'|-|z'-z|}
\eeq
Geodesics in $D$ are either diameters of the unit circle or circular arcs orthogonal to it.

%
The surface element in $D$ is given by
$$ds^2 = \frac{dz\overline{dz}}{(1-|z|^2)^2}.$$


In the rest of the paper we use the Poincar\'e disk model. This is a subjective choice essentially driven by the fact that this model exhibits in an obvious manner the rotational symmetry of the hyperbolic plane.

We now detail the relationships between SSDP(2) and its representation in the Poincaré unit disk $D$.
We also describe how the action of the direct isometries of $D$ on this representation lifts to SSDP(2). This is important since it allows us to give an interpretation
in terms of image-based operations, hence biological and computational, of the action of an isometry in $D$. This will turn out to be most important in the sequel.

A unit determinant structure tensor $\T$ is a $2 \times 2$ symmetric positive definite matrix defined by \eqref{eq:tensorabc} and satisfying $ab-c^2=1$.
This implies $a+b \geq 2$ because $a+b \geq 2\sqrt{ab}= 2\sqrt{1+c^2}$. The linear change of variables
\begin{equation}\label{eq:DH2}
 x_0=\frac{a+b}{2} \quad x_1=\frac{a-b}{2} \quad x_2=c
\end{equation}
establishes a one to one mapping from the set of structure tensors to the hyperboloid model of $H^2$ from which we deduce the correspondences with the Poincaré disk $D$. The corresponding point in $D$ is represented by the complex number
\begin{equation}\label{eq:TensortoD}
 z=\frac{1}{2+a+b}(a-b+2 i c).
\end{equation}
$z$ satisfies
\[
 0 \leq |z|=\frac{a+b-2}{a+b+2} < 1.
\]
We note ${\rm Tr}$ the trace $a+b$ of $\T$. This shows that the border of $D$, the unit circle, corresponds to the tensors such that ${\rm Tr} \to \infty$.

Conversely, given a complex number $z=z_1+iz_2$ representing a point of $D$, the corresponding tensor coordinates are given by
\begin{equation}\label{eq:DtoTensor}
 \left\{
\begin{array}{lcl}
 a & = & \frac{(1+z_1)^2+z_2^2}{1-z_1^2-z_2^2}\\
b & = & \frac{(1-z_1)^2+z_2^2}{1-z_1^2-z_2^2}\\
c & = & \frac{2z_2}{1-z_1^2-z_2^2}
\end{array}
\right.
\end{equation}
Note that equation \eqref{eq:TensortoD} is the ``Tensor to $D$ dictionary'' that allows us to translate statements about structure tensors to statements about points in the unit disk and equations \eqref{eq:DtoTensor} are the ``$D$ to Tensor'' dictionary.

Also note that we have
\[
 d_0(\mathcal{T},\mathcal{T}')=d_1(y,y')=d_2(z,z')=d_3(u,u')
\]
for all pairs $(\mathcal{T},\mathcal{T}')$ of unit determinant structure tensors represented by $(y,y')$ in the hyperboloid model, $(z,z')$ in the Poincar\'e disc model, and $(u,u')$ in the Poincaré half-plane model (see supplementary text material S2). In particular, the distance \eqref{eq:d0} defined between two structure tensors is equal to the Hyperbolic distance between their representations in the Poincaré half-plane or unit disk.

We now describe the isometries of $D$, i.e. the transformations that preserve the distance $d_2$.
Here again we recall some basic facts, now focusing on the hyperbolic geometry of the Poincar\'e disc. We refer to classical textbooks in hyperbolic geometry for details, e.g., \cite{katok:92}. 
The direct isometries (preserving the orientation) in $D$ are the elements of the special unitary group, noted ${\rm SU}(1,1)$, of $2\times 2$ Hermitian matrices with determinant equal to 1. Given\footnote{$\overline{z}$ indicates the complex conjugate of the complex number $z$.}
\[
 \gamma =\left[
\begin{array}{cc}
\alpha & \beta \\ \overline\beta & \overline\alpha \end{array}
\right] \ \text{such that}\ |\alpha|^2-|\beta|^2=1,
\]
an element of ${\rm SU}(1,1)$, the corresponding isometry $\gamma$ in $D$ is defined by
\beq \label{eq:motionD}
\gamma \cdot z = \frac{\alpha z+\beta}{\overline\beta z+\overline\alpha},~~z\in D
\eeq
Orientation reversing isometries of $D$ are obtained by composing any transformation (\ref{eq:motionD}) with the reflection $\kappa:~z\mapsto \overline z$. The full symmetry group of the Poincar\'e disc is therefore (see table \ref{table:glossary})
$${\rm U}(1,1)={\rm SU}(1,1)\cup \kappa\cdot {\rm SU}(1,1).$$
The action of the group ${\rm SU}(1,1)$ on the Poincar\'e disc $D$, is equivalent to the conjugation on the set of structure tensors. We call it the lifted action of ${\rm SU}(1,1)$ to the set of structure tensors.
Indeed, let
\[
 \gamma =\left[
\begin{array}{cc}
 \alpha & \beta\\
\overline{\beta} & \overline{\alpha}
\end{array}
\right],\,\alpha=\alpha_1+i \alpha_2,\,\beta=\beta_1+i \beta_2
\]
be an element of ${\rm SU}(1,1)$, whose action on $D$ is given by (\ref{eq:motionD}), then it can be shown by an easy computation that the  lifted action on the corresponding structure tensor $\T$ is
\begin{equation}\label{eq:action}
 \tilde{\gamma} \cdot \mathcal{T}=^t\hspace{-0.1cm}\tilde{\gamma}\hspace{.05cm} \mathcal{T} \tilde{\gamma},
\end{equation}
where
\begin{equation}\label{eq:SL2R}
 \tilde{\gamma}=\left[
\begin{array}{cc}
 \alpha_1+\beta_1 & \alpha_2+\beta_2\\
\beta_2-\alpha_2 & \alpha_1-\beta_1
\end{array}
\right]\quad \in {\rm SL}(2,\R).
\end{equation}
Equation \eqref{eq:action} is important. It shows that the ``lifted'' action on a given structure tensor $\T$ of an isometry $\gamma$ of $D$ is simply a change of coordinates $\tilde{\gamma}$ in the image plane, where the relation between $\gamma$ and $\tilde{\gamma}$ is given by equation \eqref{eq:SL2R}. We show below that these changes of coordinate systems have very simple interpretations for many of the subgroups that generate ${\rm SU}(1,1)$.

Because isometries are conformal maps, they preserve angles. However they do not transform straight lines into straight lines. Given two points $z\neq z'$ in $D$, there is a unique geodesic passing through them: the portion in $D$ of the circle containing $z$ and $z'$ and intersecting the unit circle at right angles. This circle degenerates to a straight line when the two points lie on the same diameter. Any geodesic uniquely defines the reflection through it. Reflections are orientation reversing, one representative is the complex conjugation $\kappa$ (reflection through the geodesic $\R$): $\kappa \cdot z=\overline{z}$. 

\begin{figure}[!ht]
\centerline{
\includegraphics[width=3.33cm]{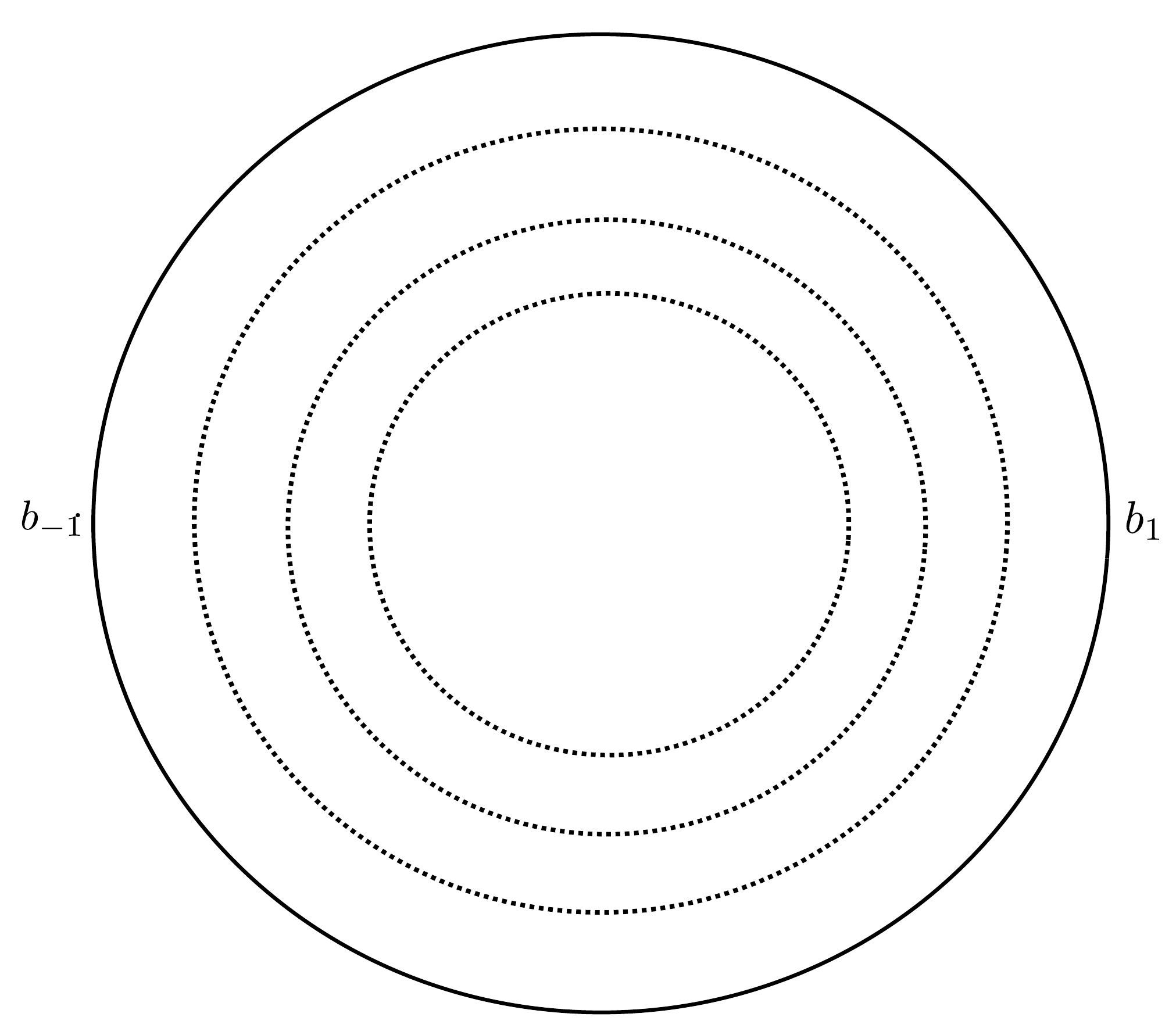}\hspace{0.01cm}\includegraphics[width=3.33cm]{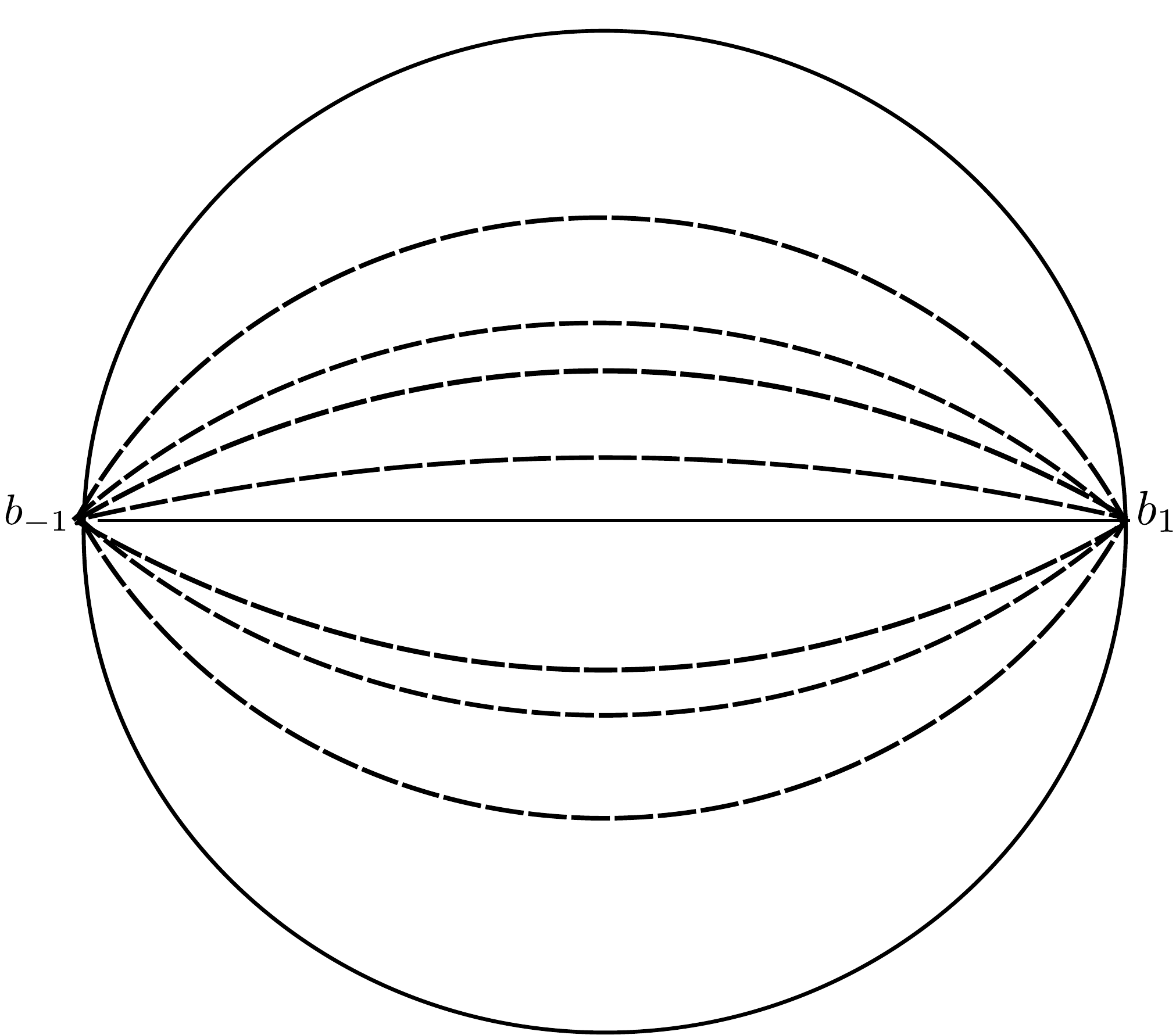}\hspace{0.01cm}\includegraphics[width=3.33cm]{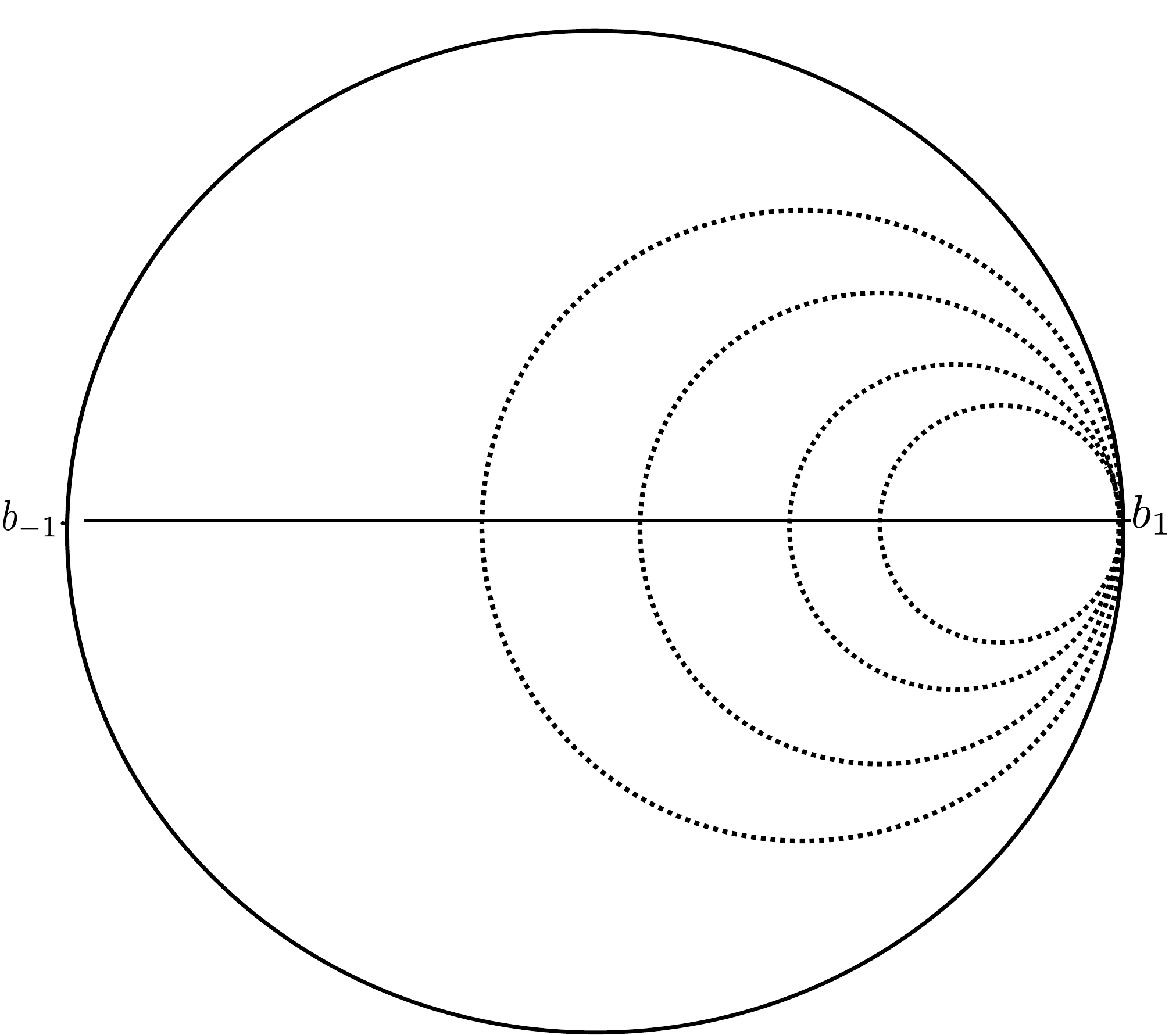}
}
\caption{{\bf The orbits in the Poincaré disk $D$ of the three groups $K$, $A$ and $N$.}}
\label{fig:orbits}
\end{figure}

Let us now describe the different kinds of direct (orientation preserving) isometries acting in $D$. Thanks to \eqref{eq:action}, they induce some interesting lifted actions on the set ${\rm SSDP}(2)$ of structure tensors that we also describe. We first define the following one-parameter subgroups of ${\rm SU}(1,1)$:
\begin{definition*} \label{def_KAN}
\[
\left\{
\begin{array}{lcl}
K & = & \{r_\varphi =\left[\begin{array}{cc} e^{i\varphi/2} & 0 \\ 0 & e^{-i\varphi/2} \end{array} \right], \quad \varphi\in S^1\}\\
&&\\
A & = & \{a_t=\left[\begin{array}{cc} \cosh t & \sinh t \\ \sinh t & \cosh t \end{array} \right], \quad t \in \R\}\\
&&\\
N & = & \{n_s=\left[\begin{array}{cc} 1+is & -is \\ is & 1-is \end{array} \right], \quad s \in \R\}
\end{array}
\right.
\]
\end{definition*}
Note that $r_\varphi\cdot z=e^{i\varphi}\, z$ for $z\in D$ and also, $a_t\cdot 0=tanh(t)$. The elements of $A$ are sometimes called ``boosts'' in the theoretical Physics literature \cite{balazs-voros:86}.
The corresponding, lifted, elements of ${\rm SL}(2,\R)$ are, according to \eqref{eq:SL2R},
\begin{equation}\label{eq:tildeKAN}
\left\{
 \begin{array}{lcl}
\tilde{r}_\varphi&=&\left[\begin{array}{cc} \cos \frac{\varphi}{2} & \sin  \frac{\varphi}{2} \\ -\sin  \frac{\varphi}{2} &  \cos \frac{\varphi}{2} \end{array} \right]\\
&&\\
\tilde{a}_t&=&\left[\begin{array}{cc} e^{t} & 0 \\ 0 & e^{-t} \end{array} \right]\\
&&\\
\tilde{n}_s &=& \left[\begin{array}{cc} 1 & 0 \\ -2s & 1 \end{array} \right],
\end{array}
\right.
\end{equation}
They generate three subgroups, noted $\tilde{K}$, $\tilde{A}$ and $\tilde{N}$, of ${\rm SL}(2,\R)$
Then the following theorem holds (Iwasawa decomposition, see \cite{iwaniec:02}).
\begin{theorem*}
\[
{\rm SU}(1,1)=KAN \quad {\rm SL}(2,\R)=\tilde{K}\tilde{A}\tilde{N}
\]
\end{theorem*}
\noindent 
This theorem allows us to decompose any isometry of $D$ as the product of at most three elements in the groups $K$, $A$ and $N$. The group $K$ is the orthogonal group O(2) which fixes the center $O$ of $D$. Its orbits are concentric circles. The orbits of $A$ converge to the same limit points of the unit circle $\partial D$ $b_{\pm 1}=\pm 1$ when $t\rightarrow\pm\infty$. They are the circular arcs in $D$ going through the points $b_1$ and $b_{-1}$. In particular the diameter $(b_{-1},b_{1})$ is an orbit. The orbits of $N$ are the circles inside $D$ and tangent to the unit circle at $b_1$. These circles are called {\em horocycles} with base point $b_1$. Because of this property,
$N$ is called the horocyclic group. These orbits are shown in figure \ref{fig:orbits}.
\noindent
Any direct isometry which is not $\pm {\rm Id}_2$ (${\rm Id}_2$ is the $2\times 2$ identity matrix) falls into one of the following three classes (representatives of which being given by elements of $K$, $A$ and $N$ respectively):
\begin{itemize}
\item[(i)] Elliptic elements: one fixed point in $D$;
\item[(ii)] Hyperbolic elements: two (and only two) fixed points on the unit circle;
\item[(iii)] Parabolic elements: one and only one fixed point on the unit circle.
\end{itemize} 

Let us interpret the meaning of \eqref{eq:action} in particular in view of the above definition of the three groups $K$, $A$, $N$ and equations \eqref{eq:tildeKAN}, i.e., what is the corresponding action on the unit determinant structure tensor $\T$ represented by the point $z$ of $D$ when $z$ is acted upon by the two groups $K$ and $A$ \footnote{There is no corresponding intuitive interpretation for the horocyclic group $N$.}.
\begin{enumerate}
	\item The action $r_\varphi \cdot z$ of an element $r_\varphi$ of $K$ on an element $z$ of $D$ lifts to the conjugation $\tilde{r}_\varphi^T \mathcal{T} \tilde{r}_\varphi$ of the structure tensor $\mathcal{T}$ represented by $z$. This is equivalent to say that we rotate by $\varphi/2$ the orthonormal basis $(\mathbf{e}_1,\mathbf{e}_2)$ in which the coordinates of $\mathcal{T}$ are expressed.
	\item The action $a_t \cdot z$ of an element $a_t$ of $A$ on an element $z$ of $D$ lifts to the conjugation $\tilde{a}_t^T \mathcal{T} \tilde{a}_t$ of the structure tensor $\mathcal{T}$ represented by $z$. This is equivalent to saying that we scale the first vector of the orthonormal basis $(\mathbf{e}_1,\mathbf{e}_2)$ in which the coordinates of $\mathcal{T}$ are expressed by $e^{t}$ and the second by $e^{-t}$.
\end{enumerate}
At this point the reader may wonder what is the biological an/or computational relevance of imposing on the structure tensors the constraint that their determinant be equal to 1. This is indeed
a somewhat unnatural assumption. In the supplementary text material S3 we propose a spherical model of the whole set, ${\rm SDP}(2)$, of structure tensors that is obtained by piecing together into a sphere the scaled Poincaré disk models of each subset of structure tensors of constant determinant. Because of this model we can restrict, without loss of generality,  our attention to the set ${\rm SSDP}(2)$ of unit determinant structure tensors.


A question which will be important in our subsequent analysis of pattern formation  is that of the periodic tilings of the hyperbolic plane, i.e., the existence of a compact domain $F$ of $D$ and of a discrete subgroup $\Gamma$ (a so-called {\em Fuchsian group} \cite{katok:92}) of the isometry group of $D$, such that\footnote{$\mathring{F}$ is the interior of the closed set $F$, i.e. the largest open set included in $F$.}
\begin{eqnarray*}
&(i)&~ \mathring{F}\cap (\gamma\cdot F)=\emptyset \rm{ ~for~all~} \gamma\in\Gamma, ~\gamma\neq Id \\
&(ii)&~D = \bigcup_{\gamma\in\Gamma} \gamma\cdot F
\end{eqnarray*}
Such an $F$ is called a {\em fundamental domain} for $\Gamma$ which is furthermore called co-compact if $F$ is compact. This property is relevant to the upcoming discussion about the eigenvalues and the eigenfunctions of the restriction to their Dirichlet regions of the Laplace-Beltrami\footnote{The Laplace-Beltrami operator is the generalisation of the Laplace operator to operate on functions defined on surfaces, or more generally on Riemannian manifolds.} operator.

This definition is similar to the one which holds for the discrete subgroups of the isometry group, noted ${\rm E}(2,\R)$, of the Euclidean plane. It is well-known that periodic tilings of the Euclidean plane are associated with lattice subgroups of the translation group $\R^2$, i.e. discrete subgroups $\Gamma$ defined by a vector basis $({\bf e}_1, {\bf e}_2)$ and $\Gamma=\{m{\bf e}_1+n{\bf e}_2,~(m,n)\in\Z^2\}$. The maximal subgroup of ${\rm O}(2)$ which leaves the lattice invariant is called the holohedry of the lattice.
If $\|{\bf e}_1\|=\|{\bf e}_2\|$, the only possibilities are when these two vectors make a right angle (square lattice, holohedry $D_4$), an angle equal to $\pi/3$ or $2\pi/3$ (hexagonal lattice, holohedry $D_6$), or an angle different from those ones (rhombic lattice, holohedry $D_2$). A ``degenerate'' case is when any period is allowed in one direction, in other words $\Gamma=\{m{\bf e}_1+y{\bf e}_2,~{\bf e}_2\perp {\bf e}_1,~(m,y)\in\Z\times\R\}$. In this case the fundamental domain is non compact and fills a ``strip'' between two parallel lines orthogonal to ${\bf e}_1$ and distant of length $\|{\bf e}_1\|$. Since the quotient $\R^2/\Z^2$ is a torus, harmonic analysis for functions which are invariant under the action of $\Gamma$ reduces to Fourier series expansion for bi-periodic functions in the plane.  

In the hyperbolic case the problem is more complex. The reason is that the Euclidean plane, which can be viewed as the symmetric space ${\rm E}(2,\R)/{\rm O}(2)$, is an Abelian group, while the Poincar\'e disc $D\simeq {\rm SU}(1,1)/{\rm SO}(2)$ is a symmetric space but has no such group property. 
It was shown by Poincar\'e in 1880 that any regular polygon\footnote{In fact, the size of the polygon is important as described in a theorem due to Poincaré \cite[Theorem 4.3.2]{katok:92}.} in $D$ generates a periodic tiling by acting recursively with reflections along the edges of the ``tiles''  \cite{katok:92}. 

Harmonic analysis for $\Gamma$-invariant functions in $D$ is difficult and relies upon the theory of modular functions and associated concepts (see \cite{gelfand-graev-etal:90,iwaniec:02}).

One special and important case for our purpose is the following. Let us consider the horocycle, noted $\xi_0$, with base point $b_1\in\partial D$ and passing through the center $O$ of $D$. Let $\xi_t$ be the image of $\xi_0$ under the hyperbolic transformation $a_t$ (see the definition above), i.e. the circle tangent to $\partial D$ at $b_1$ and going through the point $a_t \cdot O$. The map $t\in\R \mapsto a_t$ is a group homomorphism. Therefore, given $T>0$, the set $\{a_{nT},~n\in\Z\}$ is a discrete subgroup of the group $A$ whose fundamental (non compact) domain is delimited, for example, by the horocycles $\xi_0$ and $\xi_T$. This ``croissant'' shaped domain is the analogue  the ``strip'' in the Euclidean case. The ``lines'' perpendicular to the horocycles are the geodesics emanating from the point $b$. Any function in $D$ which is invariant under the action of the horocyclic group $N$ and which is ``periodic'' with respect to a subgroup of $A$ as above, can therefore be developed in Fourier series in the variable $t$. We shall come back to this later in more details.

Fundamental regions may be unnecessarily complicated, in particular they may not be connected. An alternative definition is that of a Dirichlet region of a Fuchsian group. Given two points $z$ and $z'$ of $D$ we recall that the perpendicular bisector of the geodesic segment $[z,z']$ is the unit geodesic through its midpoint (for the hyperbolic distance in $D$) orthogonal to $[z,z']$. If $z$ is a point of $D$ which is not fixed by any element of a Fuchsian subgroup $\Gamma-\{{\rm Id}\}$ of SU(1,1) (such points exist according to \cite[Lemma 2.2.5]{katok:92}) the Dirichlet region for $\Gamma$ centered at $z$ is the set noted $D_z(\Gamma)$ defined by
\[
 D_z(\Gamma)=\{p \in D\,|\,d_2(p,z) \leq d_2(p,\gamma(z)) \ \forall \gamma \in \Gamma\}
\]
It can be shown that $D_z(\Gamma)$ is a connected fundamental region for $\Gamma$, \cite[Theorem 3.2.2]{katok:92}, that generates a periodic tiling of $D$. 

We noted that the action of $\tilde{K}$ on the set of structure tensors was equivalent to a rotation of the Euclidean coordinate system. If we consider the discrete subgroup $\tilde{K}_n$ of $\tilde{K}$ (respectively $K_n$ of $K$) generated by the rotations of angles $ \pi/n$, $n \in \mathbb{N}^+$. $K_n$ is a Fuchsian group because it is obviously discrete. It is easy to find a non-compact Dirichlet region for this group showing that it is not co-compact. Nonetheless, the quotient group $D/K_n$ can be interpreted in terms of retinal properties. An element of $D/K_n$ is an equivalence class of structure tensors which are the same tensor expressed in orthonormal Euclidean coordinate systems that are rotated by multiples of $\pi/n$ with respect to each other. This makes perfect sense in terms of a discrete organisation of a visual area as an arrangement of such elements as hypercolumns at the vertixes of a periodic (Euclidean) lattice. For example, a square lattice corresponds to $n=2$ or 4, a hexagonal lattice to $n=6$.

\begin{figure}[!ht]
 \centerline{
\includegraphics[width=3.5in]{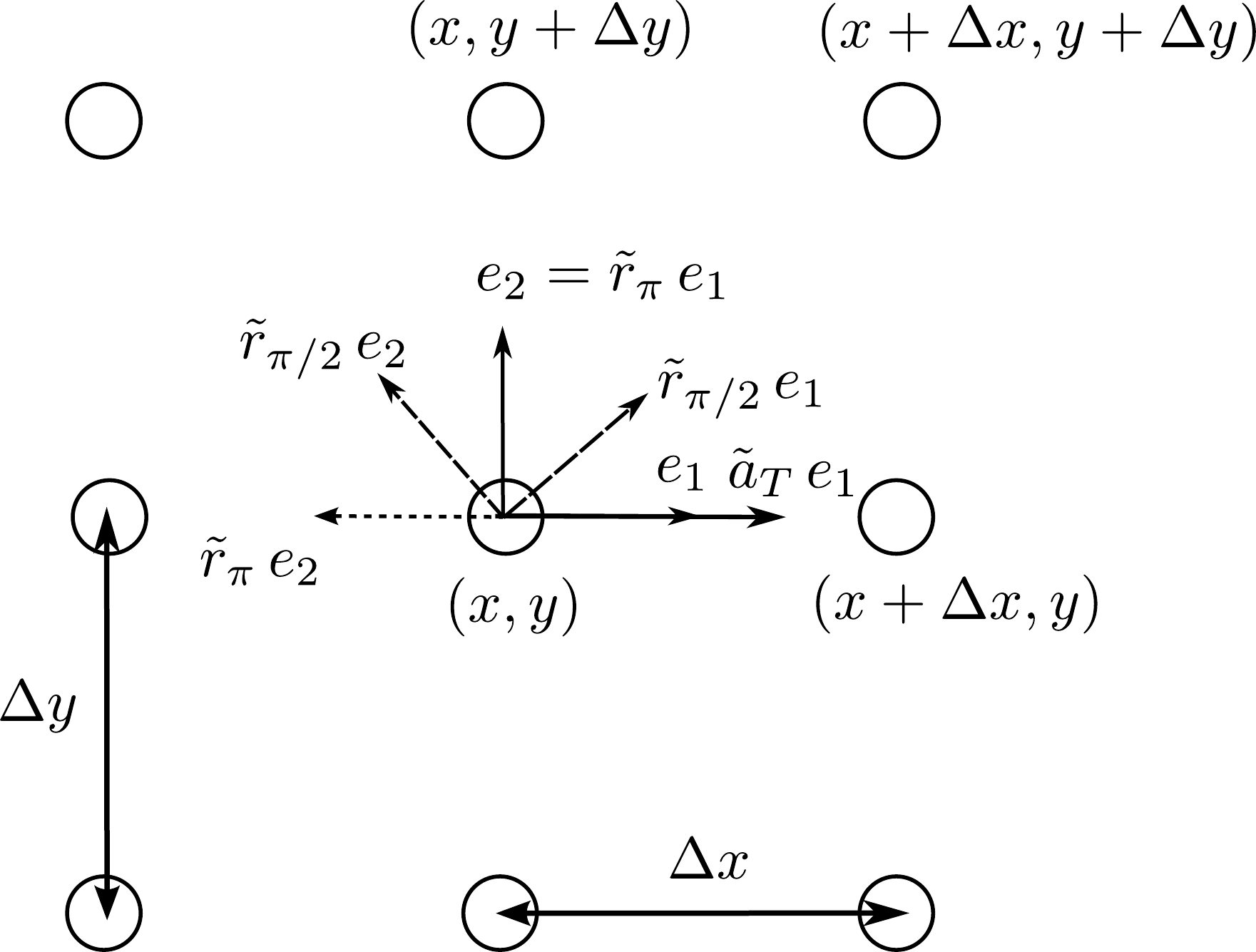}
}
\caption{{\bf The image plane: the coordinate system $(\me_1,\me_2)$ which is used to estimate the image derivatives and some of its transformations under the action of some elements of $\tilde{\Gamma}_{n,T}$ (see text).}}
\label{fig:pixels}
\end{figure}

In a similar manner, the action of $\tilde{A}$ is the multiplication of the $a$-coordinate of the tensor by $\lambda=e^{2t}$ and of the $b$-coordinate by $1/\lambda$, leaving $c$ unchanged. Remember that $a$ has the interpretation of the spatial average of the square of the spatial derivative $I_x$ of the image intensity in the $x$ direction, $b$ of the average of the square of the spatial derivative $I_y$ of the image intensity in the $y$ direction, and $c$ of the spatial average of the product $I_xI_y$, see figure \ref{fig:pixels}. $I_x$  is approximated by the cortical structure by such quantities as $(I(x+\Delta x,y)-I(x,y))/\Delta x$, and a similar expression for $I_y$ involving a distance $\Delta y$. 
This requires that the distances $\Delta x$ and $\Delta y$ be known to the neuronal elements something unlikely to happen. 
Their product $\Delta x \Delta y$ has the dimensionality of an area proportional to the average area of the tiles of the periodic (Euclidean) lattice formed by the hypercolumns. The action of $\tilde{A}$ on a structure tensor is therefore equivalent to changing $\Delta x$ and $\Delta y$ while preserving their product, the tile area. 

For a given value $T$ of the real parameter $t$ we note $A_T$ (respectively $\tilde{A}_T$) the cyclic subgroup of $A$ (respectively of $\tilde{A}$) generated by the group element $a_T$ (respectively $\tilde{a}_T$).

We consider the free product\footnote{The free product of two groups $G$ and $G'$ is the set of ``words'' composed of ``letters'' that are elements of $G$ and $G'$, see   \cite{lang:93} for details.} $\Gamma_{n,T}=K_n*A_T$ of the two groups $K_n$ and $A_T$. It is an infinite subgroup  of ${\rm SU}(1,1)$. It is generated by the elliptic element $r_{2\pi/n}$ (see equations \eqref{eq:tildeKAN}) and the hyperbolic element $a_T$. Why is this group important? If we consider the quotient group $D/\Gamma_{n,T}$ an equivalence class $\tilde{z}$ is the orbit of $z$, a  point of $D$, under the action of $\Gamma_{n,T}$ or, equivalently, an equivalence class $\overline{\T}$ of the unit determinant structure tensor $\T$ represented by $z$ under the action of the lifted subgroup $\tilde{\Gamma}_{n,T}=\tilde{K}_n * \tilde{A}_T$ of ${\rm SL}(2,\R)$. All tensors in  $\overline{\T}$ are representations of the same ``intrinsic'' tensor in coordinate systems that differ only by finite iterations of rotations of $\pi/n$ and scalings by $e^{T}$. In other words this equivalence class reflects the kind of geometric ``ignorance'' that we may expect from the neuronal populations that deal with structure tensors. Continuing the analysis, if the group $\Gamma_{n,T}$ is Fuchsian for some values of $T$ and $n$ then we are naturally led to consider one of its fundamental domains or Dirichlet regions. As mentioned above it defines a periodic tiling of $D$ which can be used to define functions in $D$ that are invariant with respect to the action of $\Gamma_{n,T}$ and hence functions of structure tensors that are invariant with respect to the action of $\tilde{\Gamma}_{n,T}$.

The question of whether $\Gamma_{n,T}$ is a Fuchsian group depends on the respective values of $T$ and $n$. The question has been answered in general for two elements of ${\rm SU}(1,1)$ \cite{rosenberger:72,gilman-maskit:91,gilman:95}. It can be cast as an algorithm whose complexity is polynomial \cite{jiang:01}. For the values of the rotation angle of interest to us we have the following proposition whose proof can be found in supplementary text material S4.
\begin{proposition*}\label{prop:Fuchsian}
 $\Gamma_{2,T}$ is a Fuchsian group for all $T \neq 0$. $\Gamma_{4,T}$ (respectively $\Gamma_{6,T}$) is a Fuchsian group if $\cosh T \geq \sqrt{2}$ (respectively
if $\cosh T \geq 2$).
\end{proposition*}
At this point we do not know whether some of these Fuchsian groups are co-compact. 

\section*{Results}
The dynamics of equation (\ref{eq:neuralmass}) depends on the input signal $I(\T,\tau)$, the sigmoid function $S$ and the connectivity function $w(\T,\T')$. 
In the Poincaré disk this equation reads, with a slight abuse of notations
\begin{equation}\label{eq:neuralmassD}
 V_\tau(z,\tau)=-\alpha V(z,\tau)+\int_D w(z,z')S(V(z',\tau))\,dm(z')+I(z,\tau),
\end{equation}
where $z$ and $z'$ are the representations of $\T$ and $\T'$ and 
\begin{equation}\label{eq:PDarea}
dm(z')=\frac{dz_1'\,dz_2'}{(1-|z'|^2)^2}\ z'=z_1'+iz_2',
\end{equation}  
is the Poincaré disk model area element.

We only consider this equation in the sequel. The reader can easily convert all the results to the set of tensors using the dictionary previously developed.

Let us assume from now on that $I=0$. This corresponds to an isolated set of neural populations, which however interact among themselves and may have non trivial states and dynamics. Our aim is to analyse this problem from the point of view of the bifurcation from a trivial state. Indeed, assuming that a solution $V$ of this equation is homogeneous, meaning that it does not depend upon the structure tensor, it follows that the equation to solve reduces to a single real equation of the form
\[
-V+W_0S(V)=0
\]
where $W_0=\int_D w(0,z')\,dm(z')$. This equation has a single solution whatever $W_0$ and $\mu>0$ (see equation \eqref{eq:sig}). We may perform a simple change of coordinates to shift this solution to $0$. This is equivalent to the choice of a sigmoid function of the form
\begin{equation}\label{eq:S0}
S_0(x)=\frac{1-e^{-\mu x}}{2(1+e^{-\mu x})}
\end{equation}
in equations \eqref{eq:neuralmass} and (\ref{eq:neuralmassD}), which we will assume in the following. A fundamental property of this new equation is that its symmetries are preserved by this change of variables. 

With these choices $V=0$ is a solution for all values of $W_0$ and $\mu$. Note that, when $\mu$ is small, this solution is dynamically stable against perturbations, at least against those which are small in $L^2$-norm. We may therefore ask what happens when $\mu$ is increased.  In order to answer this question we perform a bifurcation analysis of the solution of equation (\ref{eq:neuralmassD}) with $S=S_0$ with respect to the parameter $\mu$. 


\subsection*{Hyperbolic waves in the Poincar\'e disc}
We therefore consider equation (\ref{eq:neuralmassD}). The next step in the analysis of the bifurcations of its solutions is to look at the linearized equation and determine the critical values of the slope $\mu$ at which the trivial solution $V=0$ is destabilized under the influence of some biologically admissible (hence bounded) perturbations. For this we would like to proceed as in the Euclidean case, that is, by looking for perturbations in the form of elementary plane waves, the superposition of which defines a {\em periodic pattern} in the space $D$ (or $\R^2$ in the Euclidean case). 

Let us first recall the Euclidean setting. In this case plane waves are called planforms and have the general form $e^{i{\bf k}\cdot\mr}$ where ${\bf k}$ is any vector in $\R^2$ (the "wave vector"). Each planform is an eigenfunction of the Laplace operator $\Delta$ corresponding to a real eigenvalue\footnote{$\|\mathbf{k}\|$ is the Euclidean norm of the vector $\mathbf{k}$.}: 
$$
\Delta e^{i{\bf k}\cdot\mr}=-\|{\bf k}\|^2e^{i{\bf k}\cdot\mr},\,\mr \in \R^2.
$$
The fact that the eigenvalue does not depend upon the direction of the wave vector reflect the rotational invariance of the Laplace operator. Moreover, a given planform $e^{i\mk \cdot\mr}$ is clearly invariant under translations in $\R^2$ by any vector $\me$ satisfying the condition $\mk \cdot\me=2n\pi$ where $n\in\Z$ (it clearly does not depend upon the coordinate along the axis orthogonal to $\mk$). It is an elementary but fundamental fact of Euclidean geometry that given any two vectors $\mk_1$, $\mk_2$ of equal length, we can  define the periodic lattice ${\cal L}$ spanned in the plane by ${\bf e}_1$ and ${\bf e}_2$ such that $\mk_i\cdot {\bf e}_j=2\pi\delta_{ij}$, and that any smooth function in the plane which is invariant under translations in ${\cal L}$ can be expanded in a Fourier series of planewaves $e^{i(m\mk_1+n\mk_2)\cdot\mr}$, $m,n\in\Z$. Therefore in a suitable space of lattice periodic functions the spectrum of the Laplace operator is discrete with real eigenvalues of finite multiplicities, the corresponding eigenfunctions  being planforms, and we can proceed to classical bifurcation analysis if the equations do not have additional degeneracies or singularities (this was the approach of  \cite{bressloff-cowan-etal:02} for the analysis of visual hallucinations formation in the cortex).

Our aim is to apply similar ideas to the case when the problem is defined in the  Poincar\'e disc instead of the Euclidean plane. A first remark is that we cannot define a periodic lattice in $D$ by just assigning two basic wave vectors ($D$ is not a vector space). There exist however a large number of periodic lattices in $D$. Those are defined by discrete subgroups of ${\rm SU}(1,1)$, and there are many such groups (called Fuchsian groups, see above). We may therefore consider functions which are invariant under the action of a Fuchsian group. Thanks to their invariance under the action of ${\rm U}(1,1)$ we know that our equations can be restricted to such functions. Moreover, if the fundamental domain of a Fuchsian group is compact (see above), it is known that the Laplace-Beltrami operator restricted to this class of functions has a discrete spectrum of real eigenvalues with finite multiplicities. However before we go further in this direction, we first need to analyze the effect of perturbations in the form of elementary waves, the hyperbolic counterpart of planforms.   

\begin{figure}[!ht]
\centerline{
\includegraphics[width=4in]{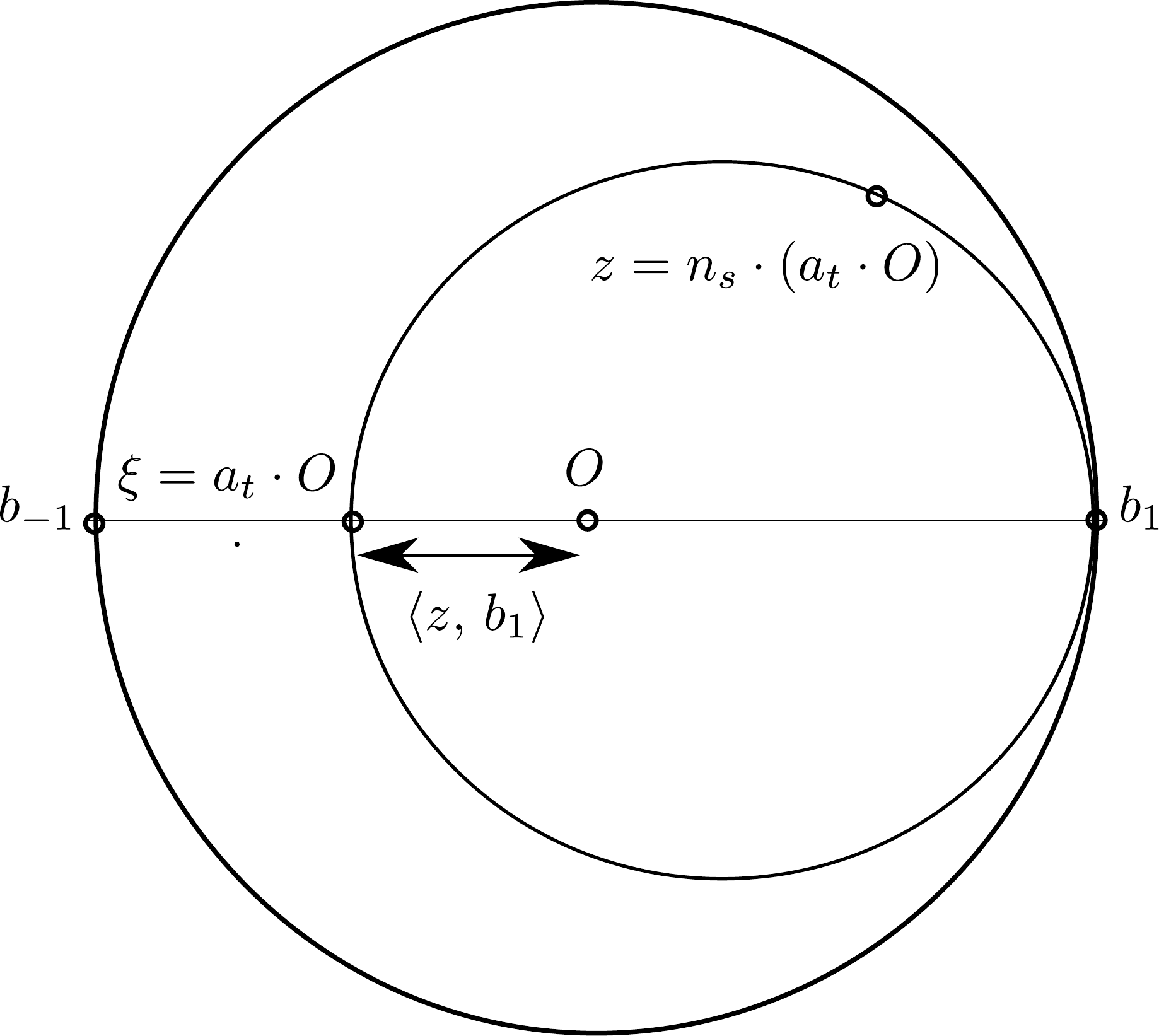}
}
\caption{{\bf The horocyclic coordinates of the point $z$ of $D$ are the real values $s$ and $t$ such that $z = n_sa_t \cdot O$.} The horocycle through $z$ is the circle tangent to $\partial D$ at $b_1$ and going through $z$.
$\langle z,b_1 \rangle$ is equal to the (hyperbolic) signed distance $d_2(O,a_t \cdot O)$ between the origin $O$ and the point $a_t \cdot O$ which is equal to $t$ and is negative if $O$ is inside the circle of diameter $(O,a_t \cdot O)$ and positive otherwise.}
\label{fig:horocyclic_coordinates}
\end{figure}

\begin{figure}[!ht]
   \centering
     \includegraphics[height=4in]{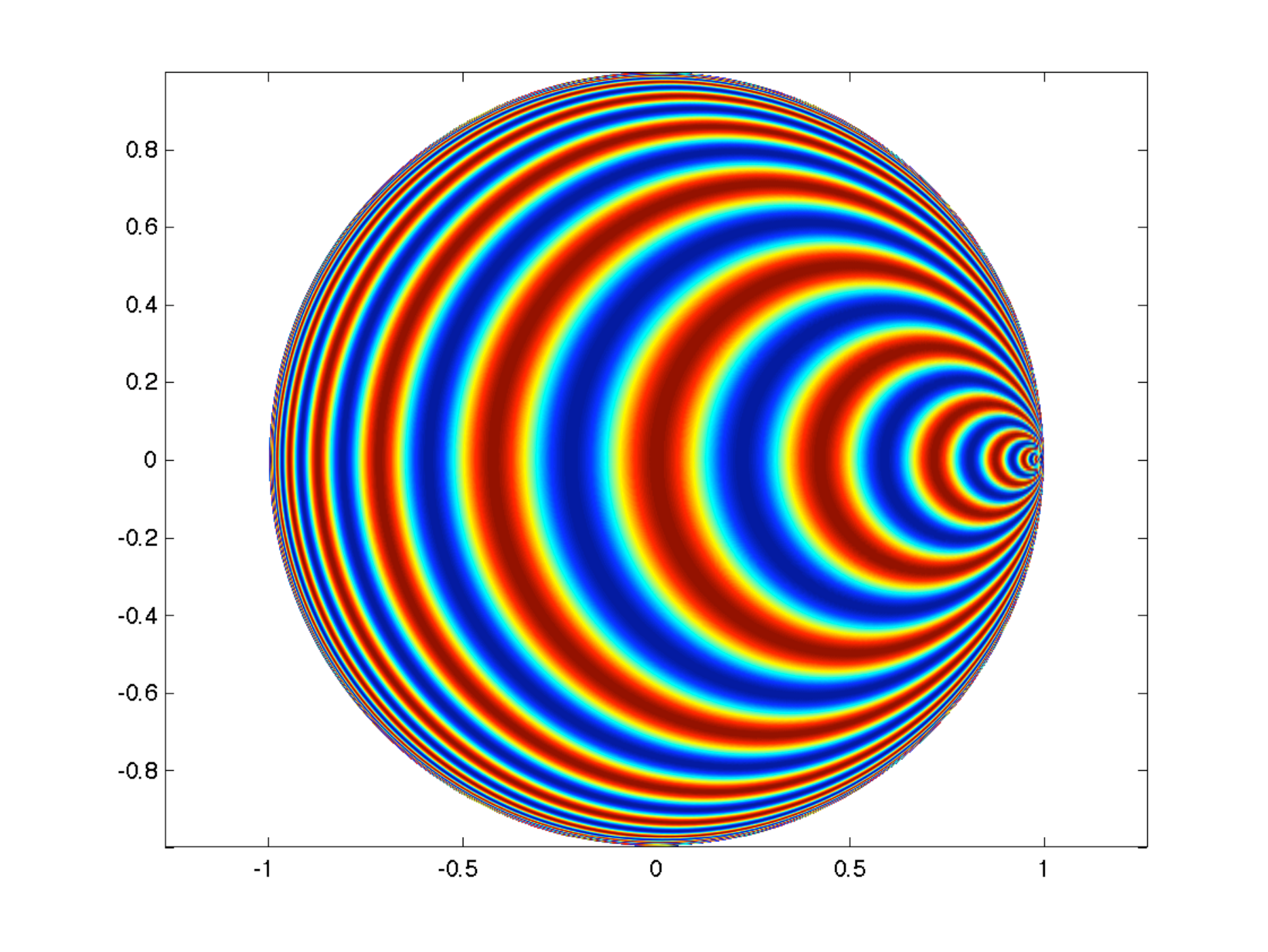}
    \caption{{\bf A periodic H-planform $e_{\alpha+i,b_1}$.} The color represents the value of the magnitude of $e_{\alpha+i,b_1}(z)$ for z varying in $D$. The periodicity is to be understood in terms of the hyperbolic distance $d_2$. The hyperbolic distance between two consecutive points of intersection of the, say yellow, circles with the horizontal axis is the same. It does not look so to our ``Euclidean'' eyes and the distances look shorter when these points get closer to the point $b_1$ on the right and to the point $b_{-1}$ of $\partial D$ on the left. These points are actually at an infinite distance from the center $O$ of $D$.}
    \label{H-planform}
 \end{figure}

Such {\em hyperbolic plane waves} have been introduced by Helgason \cite{helgason:00} and are defined as follows: Let $b$ be a point on the circle $\partial D$, which we may  take equal to $b_1=1$ by a suitable rotation. For $z\in D$, we define the "inner product" $\langle z,b \rangle$ to be the algebraic distance to the origin of the (unique) horocycle based at $b$ going through $z$. This distance is defined as the hyperbolic (algebraic) length of the segment $O\xi$ where $\xi$ is the intersection point of the horocycle and the line (geodesic) $Ob$, see figure \ref{fig:horocyclic_coordinates}. Note that $\langle z,b \rangle$ does not depend on the position of $z$ on the horocycle. In other words, $\langle z,b \rangle$ is invariant under the action of the one-parameter group $N$ (see definition above). One can check that the functions 
\[
 e_{\lambda,b}(z)=e^{(i\lambda+1)\langle z,b \rangle}, \,\lambda\in\C,
\]
are eigenfunctions of the Laplace-Beltrami operator $\Delta$ in $D$ with eigenvalues $-\lambda^2-1$.  Helgason \cite{helgason:00} used these functions to define the Fourier transform in $D$ pretty much like the elementary functions $e^{i\lambda {\bf x}\cdot\boldsymbol{\omega}}$, ${\bf x},\,\boldsymbol{\omega}\,\in\R^2$, $\|\boldsymbol{\omega}\|=1$, are used to define the usual Fourier transform in the plane.
We now define the {\em Helgason hyperbolic planforms} (or {\em H-planforms}) as the functions $e_{\lambda,b}$ with $\lambda\in\R$ or $\lambda=\alpha+i$, $\alpha\in\R$. The first case corresponds to a real eigenvalue of $\Delta$. In the second case, the eigenvalue is complex and equal to $-\alpha^2 - 2i\alpha$. The reasons for introduction of these H-planforms will become clear from the following properties: 
\begin{itemize}
\item[(i)] they are by construction invariant under the action of the subgroup $N$ (i.e. along the horocycles of base point $b_1$). They correspond therefore to wavy patterns along the geodesics emanating from $b_1$. These geodesics are parallel to each other and orthogonal to the horocycles. In that sense, these patterns are hyperbolic counterparts of the Euclidean planforms which correspond to trains of waves orthogonal to parallel straight lines in the plane (geodesics for the Euclidean metric). 
\item[(ii)] Let us express $z\in D$ in "horocylic" coordinates: $z = n_sa_t \cdot O$, where $n_s$ are the (parabolic) transformations associated with the group $N$ ($s\in\R$) and $a_t$ are the (hyperbolic) transformations associated with the subgroup $A$ ($t\in\R$), see definition above and figure \ref{fig:horocyclic_coordinates}.

It is readily seen from the definitions and formula (\ref{d2-bis}) that $\langle n_sa_t\cdot O,b_1 \rangle=t$. Therefore, in these coordinates, the H-planforms with base point $b_1$ read $e_{\lambda,b_1}(z)=e^{(i\lambda+1)t}$. 
In particular if $\lambda=\alpha+i$, then $e_{\alpha+i,b_1}$ is periodic with respect to the coordinate $t$ with period $2\pi/\alpha$. Of course the same property holds at any base point $b$ by simply rotating the planform by the angle $(b_1,b)$. The H-planform is said to be periodic in this case.
Figure \ref{H-planform} shows the pattern of a periodic H-planform. If $\lambda\in\R$, the eigenfunction $e_{\lambda,b_1}$ is not periodic due to the factor $e^t$ in front of $e^{i\lambda t}$. It does however correspond to a physically relevant wavy pattern in the sense that its "energy density" is expressed as $e_{\lambda,b_1}(t)e_{-\lambda,b_1}(t)e^{-2t}dt = dt$ and is therefore bounded (here we applied the expression $e^{-2t}dtds$ for the surface element in horocyclic coordinates, see \cite{helgason:00}).
\end{itemize}
We now proceed with the linear step of our bifurcation analysis.

\subsection*{The eigenvalue problem for equation (\ref{eq:neuralmassD})}
The linearisation of  equation (\ref{eq:neuralmassD}) at the trivial solution $V=0$, with no input and with $V:D\times\R\to \R$, reads
\begin{equation}
\label{eq:neuralmassinD}
 V_\gt(z,\gt)=-V(z,\gt)+\mu\int_D w(z,z')V(z',\gt)dm(z')
\end{equation}
where $\mu=S_0'(0)$ and $dm(z')$ is the "hyperbolic" measure in $D$ defined in equation \eqref{eq:PDarea}. Since equation \eqref{eq:neuralmassD} is invariant with respect to the isometries of $D$, we can look for solutions which are invariant under the action of the subgroup $N$. It is then appropriate to express $z,\,z'\in D$ in horocyclic coordinates:~ $z=n_sa_t\cdot O$, $z'=n_{s'}a_{t'} \cdot O$. The hyperbolic surface element  in these coordinates is expressed as \cite{helgason:00}
\begin{equation}\label{eq:surface}
 dm(z')=e^{-2t'}\,dt'\,ds'
\end{equation}
The invariance then reads
\begin{equation}
V(n_sa_t\cdot O)=V(a_t\cdot O),~\mbox{for all~}(s,t)\in\R^2
\end{equation}
The integral term in (\ref{eq:neuralmassinD}) defines a linear operator, noted $L$, on the set of average membrane potential functions $V$, which can be expressed as follows (the last identity following from the change of variable $s'-s=xe^{2t'}$ and the relation $a_tn_x=n_{xe^{2t}}a_t$ \cite{helgason:00}): 
\begin{eqnarray*}
(L\cdot V)(n_sa_t\cdot O)&=&\int_\R\int_\R w(n_sa_t\cdot O,n_{s'}a_{t'}\cdot O)V(a_{t'}\cdot O)ds'e^{-2t'}dt' \\
&=& \int_\R\int_\R w(a_t\cdot O,n_{s'-s}a_{t'}\cdot O)V(a_{t'}\cdot O)ds'e^{-2t'}dt' \\
&=& \int_\R\left(\int_\R w(a_{t-t'}\cdot O,n_x\cdot O)dx\right)V(a_{t'}\cdot O)dt'
\end{eqnarray*}
This shows that $L\cdot V$ does not depend on the coordinate $s$ (as expected).

We have reduced the problem to an integro-differential equation in the single coordinate $t$. 
Moreover, if we define 
$$\widetilde{w}(\xi)= \int_\R w(a_\xi\cdot O,n_x\cdot O)dx$$
and assume that the integral is convergent for $\xi\in \R$ (this is the case with $w$ defined by the function $g$ in (\ref{fonction:g})), then  equation (\ref{eq:neuralmassinD}) leads to the eigenvalue problem
\begin{equation} \label{ev_equation}
\sigma \widetilde{V} = -\widetilde{V} + \mu\widetilde{w}\star\widetilde{V}
\end{equation}
where $\star$ is a convolution product and we have set $\widetilde{V}(t)=V(a_t\cdot O)$. This problem can be solved by applying the Fourier transform in $D$ which is defined as (see \cite{helgason:00}):
$$
\hat h(\lambda,b) = \int_D{h(z)e^{(-i\lambda +1)\langle z,b\rangle}dm(z)} 
$$
for a function $h: D \to \C$ such that this integral is well-defined.
Thanks to the rotational invariance we can restrict ourselves to the case $b=b_1=1$, which gives, in horocyclic coordinates:
\begin{equation}
\hat h(\lambda,b_1) =  \int_\R\int_\R{h(n_sa_t\cdot O)e^{(-i\lambda -1)t} dtds}
\end{equation}
Rotational invariance implies that the same equations would be obtained if an H-planform with another base point $b$ were chosen. This can be seen directly on the expression of H-planforms from the relation (see \cite{helgason:00})
$$e_{\lambda,b}(z)=e_{\lambda, r_\varphi \cdot b}( r_\varphi \cdot z),~~r_\varphi \in K,~z\in D.$$
It follows that for a given $\lambda$ and eigenvalue $\sigma$, there is in fact a full "circle" of eigenfunctions  $e_{\lambda,b}$, $b\in \partial D$.
 
 \subsection*{Bifurcation of periodic H-planforms}
 
 We assume $\lambda=\alpha+i$ in this section. This means that we are looking for solutions of (\ref{ev_equation}) of the form $e^{\sigma\gt}\,e_{\alpha+i,b_1}(z)=e^{\sigma\gt}\,e^{i\alpha t}$, $\alpha\in\R$. The H-planforms are not only invariant along horocycles, but also $2\pi/\alpha$ periodic with respect to the coordinate $t$ as shown above. If a bifurcation occurs with such a planform,  the corresponding solutions of equation (\ref{eq:neuralmassD}) will be $s$-invariant and $t$-periodic. We first look at the critical eigenvalue problem for such H-planforms.
 
Applying the Fourier transform to (\ref{ev_equation}) leads to the following expression for the eigenvalues:
\begin{equation} \label{eq:vp}
\sigma(\alpha) = -1 + \mu\hat{w}(\alpha)
\end{equation}  
where $\hat{w}$ is the Fourier transform of $\widetilde{w}$. Numerical calculation has been performed to compute $\hat{w}$ in the case when $w$ is defined by the "Mexican hat" $g$ given in (\ref{fonction:g})).
Note that the function $\widetilde{w}$ is not even (hence the operator $L$ is not symmetric). The following two properties of $\hat{w}$ are therefore not surprising \footnote{They would be false if the system were defined in the Euclidean plane instead of the Poincar\'e disc, because in this case $L$ would be a symmetric operator.}:
(i) the eigenvalues are complex in general, (ii) the graph of $\hat{w}$ shows maxima {\em and} minima. Figure \ref{figure:graphe_w^_periodicwaves} below shows the graph obtained with $\sigma_1=0.9$, $\sigma_2=1$, $\theta=0.6$, and $f:x \to x^2$ in equation \eqref{fonction:g}. 
\begin{figure}[tb]
   \centering
     \includegraphics[width=4in]{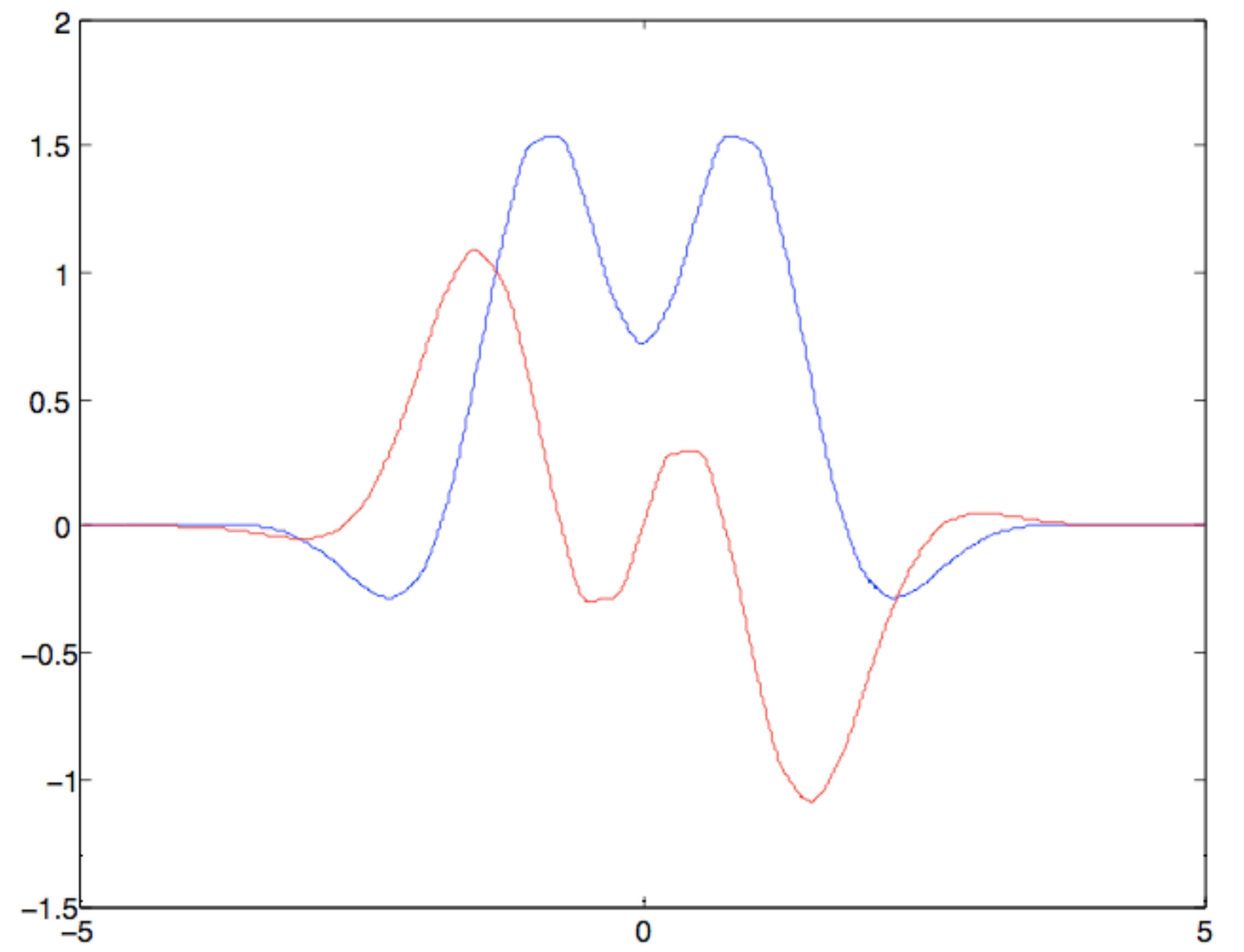}
    \caption{{\bf Real (blue) and imaginary (red) parts of $\hat{w}$ defined in equation \eqref{eq:vp} for $e_{\alpha+i,\,b}$ H-planforms, $\alpha \in \R$, see text.} We chose $\sigma_1=0.9$, $\sigma_2=1$, $\theta=0.6$ and $f(x)=x^2$ in equation \eqref{fonction:g}.}
    \label{figure:graphe_w^_periodicwaves}
 \end{figure}
 
All eigenvalues come in pairs of complex conjugates and of course $\hat{w}(-\alpha)=\overline{\hat{w}(\alpha)}$. The most unstable eigenvalues are those corresponding to the maximum of ${\rm Re}(\hat{w})$, that is, in the case of Figure \ref{figure:graphe_w^_periodicwaves}, with $|\alpha|=\alpha_c\approx 0.76$. The critical value $\mu_c$ of $\mu$ is obtained by setting the real part of $-1+\mu \hat{w}(\alpha_c)$ equal to 0. The corresponding critical eigenvalues are $\pm i\omega_0$ with $\omega_0=\mu_c {\rm Im}(\hat w(\alpha_c))$ (with the parameter values of Figure \ref{figure:graphe_w^_periodicwaves}, $\omega_0\approx 0.04$ and  $\mu=\mu_c\approx 0.65$). When $\mu<\mu_c$, small fluctuations around the trivial state of equation (\ref{eq:neuralmassD}) are damped, while as $\mu$ crosses the critical value, perturbations with period $2\pi/\alpha_c$ will grow. In fact a continuum of wave numbers close to $\alpha_c$ may also give rise to unstable modes, however we now restrict our analysis to functions which are $T$-periodic in $t$ with period $T=2\pi/\alpha_c$.
This allows us to reduce the problem to an equation bearing on functions $U$ of the time $\tau$ and the single variable $t$, which are square integrable in the interval of periodicity  $[0,T]$.    
  
It follows that a Hopf bifurcation occurs from the trivial state of equation (\ref{eq:neuralmass}) at $\mu=\mu_c$. Applying a procedure which is classical in the Euclidean case \cite{chossat-lauterbach:00}, we formulate the problem in operator terms as follows. Let $\rho=\mu-\mu_c$ be close to 0, then 
\begin{equation} \label{eq:operateurs}
\frac{dU}{d\tau} = L_0 \cdot U+\rho L_1 \cdot U+C(U)+R(U,\rho)
\end{equation}
where the operators $L_0$, $L_1$ and $NL$ are defined as follows
\begin{eqnarray*}
L_0 \cdot U &=& -U+\mu_c\widetilde{w}\star U \\
L_1 \cdot U &=& \widetilde{w}\star U \\
C(U) &=& \frac{\mu_c}{12}\widetilde{w}\star U^3,
\end{eqnarray*}
$U^3$ is the function $(t,\tau) \to (U(t,\tau))^3$, and $R(U,\rho)$ stands for the higher order terms in $U$ and $\rho$. These operators are defined in the Hilbert space $\F$ of square integrable, $\frac{2\pi}{\alpha_c}$-periodic functions $\F=L^2(\R/({\frac{2\pi}{\alpha_c}}\Z),\R)$. $L_0$ and $L_1$ are compact operators in $\F$ and $NL, R \in C^\infty(\F,\F)$. The critical eigenvalues $\pm i\omega_0$ of $L_0$ are simple. It follows from general Hopf bifurcation theory \cite{iooss-adelmeyer:98} that a branch of periodic solutions bifurcates from the trivial state at $\mu=\mu_c$, i.e at $\rho=0$, with a period $2\pi/\omega$ where $\omega$ is close to $\omega_0$, and the leading order of which has the form
$$
U_0(\tau)=\varepsilon\left[e^{i(\omega \tau+\varphi)}e_{\alpha_c+i,b_1}+e^{-i(\omega \tau+\varphi)}\overline{e_{\alpha_c+i,b_1}}\right] 
$$
where $\varphi$ is an (arbitrary) phase.
Plugging this into equation (\ref{eq:operateurs}) and passing in Fourier space at the value $\alpha=\alpha_c$ we obtain the bifurcation equation
$$
i\omega=i\omega_0+\hat w(\alpha_c)\varepsilon\rho - \frac{1}{4}\hat w(\alpha_c)\varepsilon^3 + h.o.t.
$$
from which it follows that
$$
\varepsilon = \pm 2\sqrt{\rho} + O(\rho)
$$
and $\omega-\omega_0$ is readily deduced from this by taking the imaginary part of the bifurcation equation.
The branching is therefore supercritical (for $\rho>0$) and the bifurcated, periodic solutions are stable against perturbative modes which respect the symmetries of the solutions ("exchange of stability principle", \cite{iooss-joseph:90}). At this stage however, no general stability statement can be made. 

One last remark should be made about these periodic solutions. In a suitable space of time-periodic functions (as chosen to perform the Hopf bifurcation analysis, see \cite{chossat-lauterbach:00}) the invariance under time translations of the problem induces a "temporal" symmetry by the action of the group $S^1=\R/\Z$. This group simply acts by time shifts mod $2\pi/\omega$ (the time period of the bifurcated solutions). On the other hand, another copy of $S^1$ acts on (\ref{eq:operateurs}) by shifts along the $t$ coordinate mod $2\pi/\alpha_c$ ("spatial" periodicity). These two groups act as follows on the leading term $e^{i(\omega\tau+\alpha_ct)}$ of the bifurcated solutions: 
$$(\varphi,\psi)\in S^1\times S^1\mapsto e^{i(\omega\tau+\varphi+\alpha_ct+\psi)}$$
Therefore this term, which is also the complex eigenmode for the linear part of the equation, is fixed under the action of the one-parameter subgroup of $S^1$ defined by setting $\varphi=-\psi$. By the general theory of Hopf bifurcations with symmetry \cite{golubitsky-stewart-etal:88}), this property propagates to the full solutions of (\ref{eq:operateurs}). The interpretation is that, for an observer moving along the $t$ coordinate with velocity $-\omega/\alpha_c$, the solution looks stationnary. Solutions which have this property are called relative equilibria \cite{field:96},\cite{chossat-lauterbach:00}, and in the present case they can also be named {\em H-traveling waves}. These solutions resemble a train of H-planforms propagating from the "source" at infinity which is the tangency point $b$ of the horocycles, see movie in supplementary material.

\subsection*{Bifurcation of periodic patterns in $D$} 
In the previous section we found bifurcated solutions which were periodic along the geodesics emanating from a point at infinity (i.e. on $\partial D$) and invariant along the orthogonal direction (that is, along the horocycles). This pattern corresponds to the Euclidean "strip" or "roll" pattern, with the noticeable difference that the latter are usually steady, while in our case they are uniformely traveling from the source at infinity. Is it possible to go further in the analogy with the Euclidean case? Is it possible to find bifurcating patterns which are invariant with respect to a periodic lattice (or "tesselation") in $D$, in other words patterns which are invariant under the action of a discrete subgroup $\Gamma$ of ${\rm U}(1,1)$ with a compact fundamental domain. This would be of physical relevance because it would correspond to bounded states. Moreover periodic tilings with certain types of compact "tiles" related for example to the groups $\Gamma_{n,T}$ may be specially relevant to our problem as described above.

However the occurence of such groups and the requirement of compactness of their fundamental domain obeys very strict rules. In particular, an important difference with the Euclidean tilings is that fundamental polygons for a given group have a fixed area: applying some rescaling to the domain will in general destroy the tiling property.

In any case, it results from general spectral theory on the hyperbolic plane that the spectrum of the Laplace-Beltrami operator restricted to $\Gamma$-invariant eigenfunctions, $\Gamma$ with a compact fundamental domain, is discrete and its eigenvalues have finite multiplicity \cite{gelfand-graev-etal:90,iwaniec:02}. Any smooth (square integrable) $\Gamma$-invariant function (or "automorphic function") in $D$ can be expanded in a series of eigenfunctions of $\Delta$. These eigenfunctions can be expressed in terms of $e_{\lambda,\,b}$ H-planforms ($\lambda \in \R$) as follows:
$$
\Psi_{\lambda}(z) = \int_{\partial D}{e^{(i\lambda+1)\langle z,b\rangle}dT(b)}
$$ 
where $T$ is a distribution defined on the boundary $\partial D$ of the unit disc $D$ which in addition satisfies certain equivariance relations with respect to the action of $\Gamma$ on $\partial D$. Here $\Psi_{\lambda}$ is an eigenfunction for the eigenvalue $-\lambda^2-1$, but the values of $\lambda$ depend on $\Gamma$ and there is no known simple or explicit way to compute these values and the corresponding distribution $T$.  

\begin{figure}[tb]
   \centering
     \includegraphics[width=4in]{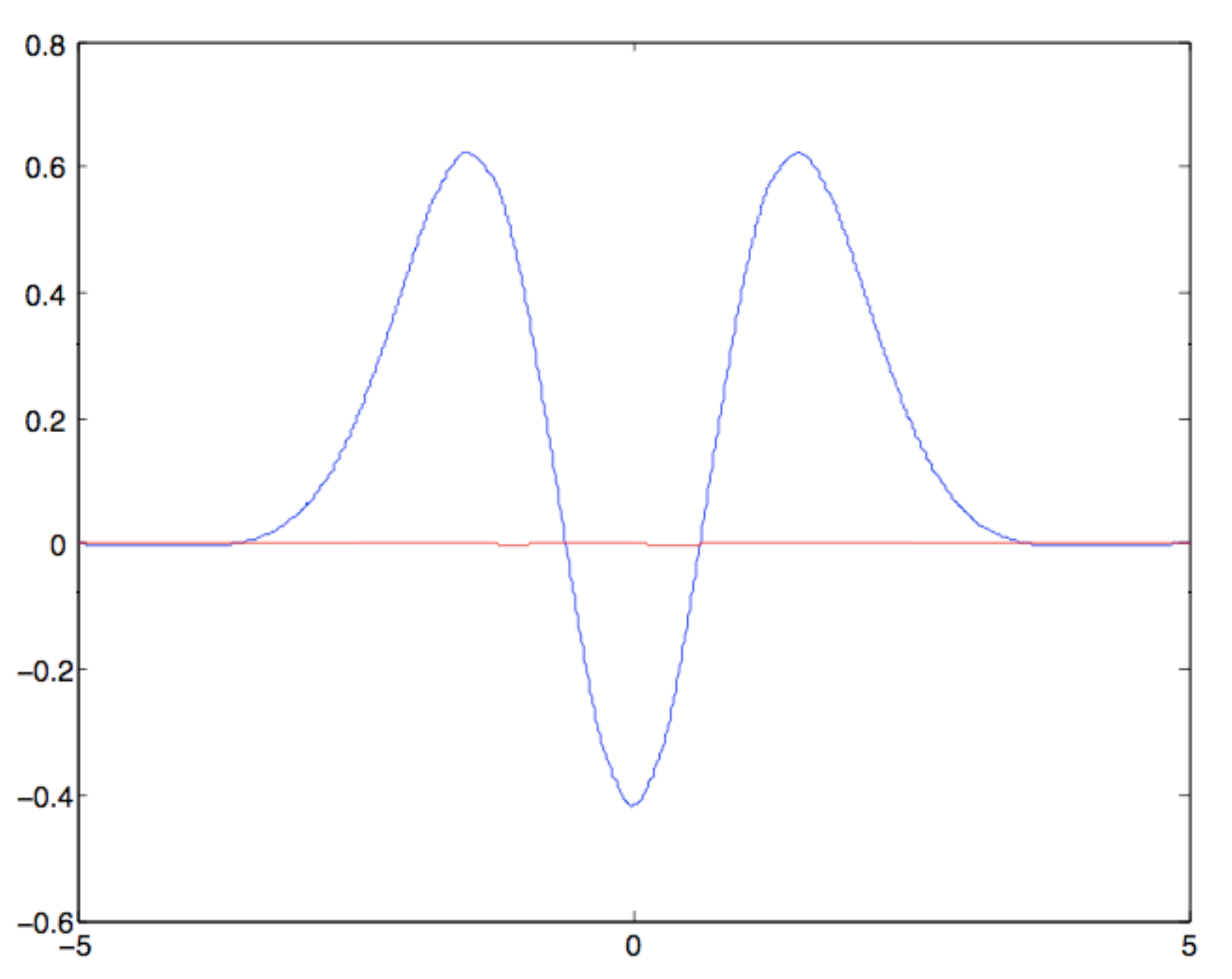}
    \caption{{\bf Real (blue) and imaginary (red) parts of $\hat{w}$ defined in equation \eqref{eq:vp} for $e_{\lambda,\,b}$ H-planforms, $\lambda \in \R$, see text.} We chose $\sigma_1=0.8$, $\sigma_2=1$, $\theta=0.7$  and $f(x)=x^2$ and $f(x)=x^2$ in equation \eqref{fonction:g}.}
    \label{figure:w^_steadypattern}
 \end{figure}

We can nevertheless determine the threshold at which perturbations along the elementary H-planforms $e_{\lambda,\,b}$ will lead to instability of the trivial state for equation (\ref{eq:neuralmassinD}). The method is completely similar to the one for periodic H-planforms. The eigenvalues are given by equation (\ref{eq:vp}). Figure \ref{figure:w^_steadypattern} shows an example of the function $\hat w(\lambda)$. As expected it takes only real values corresponding to the fact that the eigenvalues are real in this case. The most unstable eigenvalue corresponds to the maximum of the blue curve, the corresponding abscissa being the "critical" wave number $\lambda_c$. The critical value of the parameter $\mu$ is then defined by the relation $0=-1+\mu_c\hat w(\lambda_c)$, for which all eigenvalues are negative but one, the {\em critical eigenvalue}, which is at 0. Therefore when $\mu$ crosses this threshold the system undergoes a steady-state bifurcation.
The next question is to look for discrete groups $\Gamma$ such that this critical value also corresponds to $\Gamma$ invariant eigenfunctions. We have not carried out this program yet.

\begin{figure}[tb]
   \centering
     \includegraphics[width=4in]{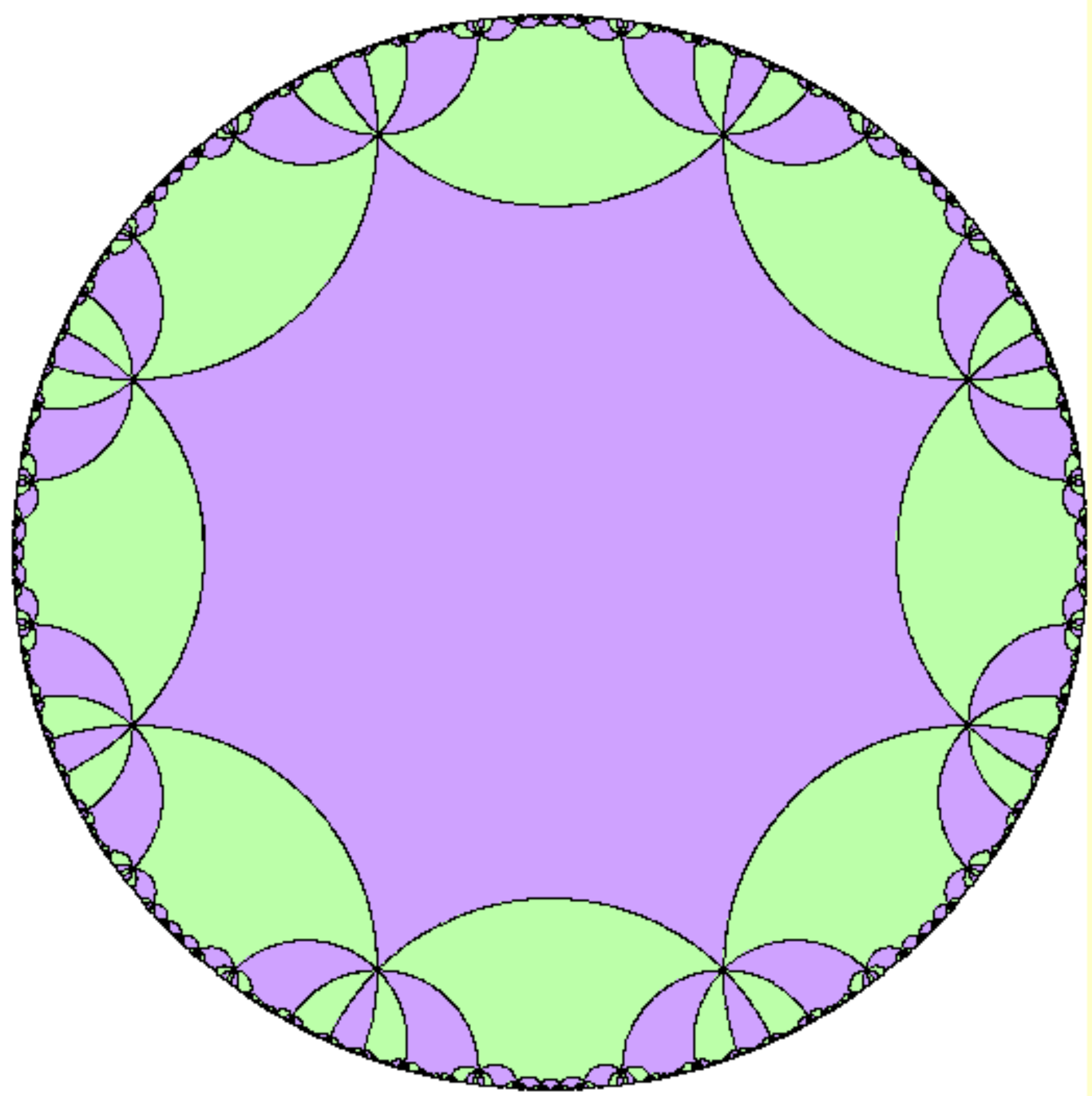}
    \caption{{\bf The fundamental domain of the octagonal Fuchsian group and the tesselation of $D$ it generates.} Two adjacent octagons are colored in different hues.}
    \label{octogone}
 \end{figure}

The computation of the eigenvalues and $\Gamma$ invariant eigenfunctions can only be achieved by numerical approximation. Only a few cases have been investigated in detail, for example the case when $\Gamma$ is the octagonal Fuchsian group (see \cite{bachelot-motet:09,balazs-voros:86}). This group, which we note $\Gamma_8$, is spanned by four "boosts" (hyperbolic elements of ${\rm SU}(1,1)$) $g_k$ with $g_0= \left[\begin{array}{cc} 1+\sqrt{2} & \sqrt{2+2\sqrt{2}} \\ \sqrt{2+2\sqrt{2}} & 1+\sqrt{2} \end{array} \right]$ and $g_k=r_{\frac{k\pi}{4}}g_0r_{-\frac{k\pi}{4}}$, $k=1,2,3$. Its fundamental domain is the regular octagon which can define a tesselation of $D$, as shown in Figure \ref{octogone}.

\begin{figure}
\centerline{
\includegraphics[width=0.75\textwidth]{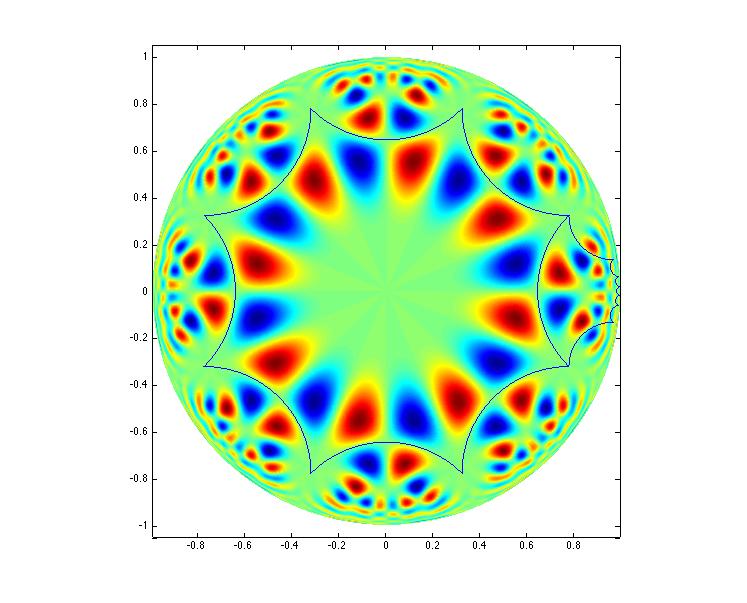}
}
\caption{\bf An example of an H-planform that is invariant with respect to the octagonal Fuchsian group. We have superimposed two fundamental domains: in the center the ``main'' one containing the origin, to its right another fundamental domain that shows the Euclidean distorsion due to the increase in the hyperbolic distance. In effect these two octagons can be exactly superimposed through the action of a hyperbolic isometry. The color encodes the value of the H-planform, blue indicates negative values, red indicate positive values, green indicates values close to 0.}
\label{fig:image2_octogone}
\end{figure}

\begin{figure}
\centerline{
\includegraphics[width=0.75\textwidth]{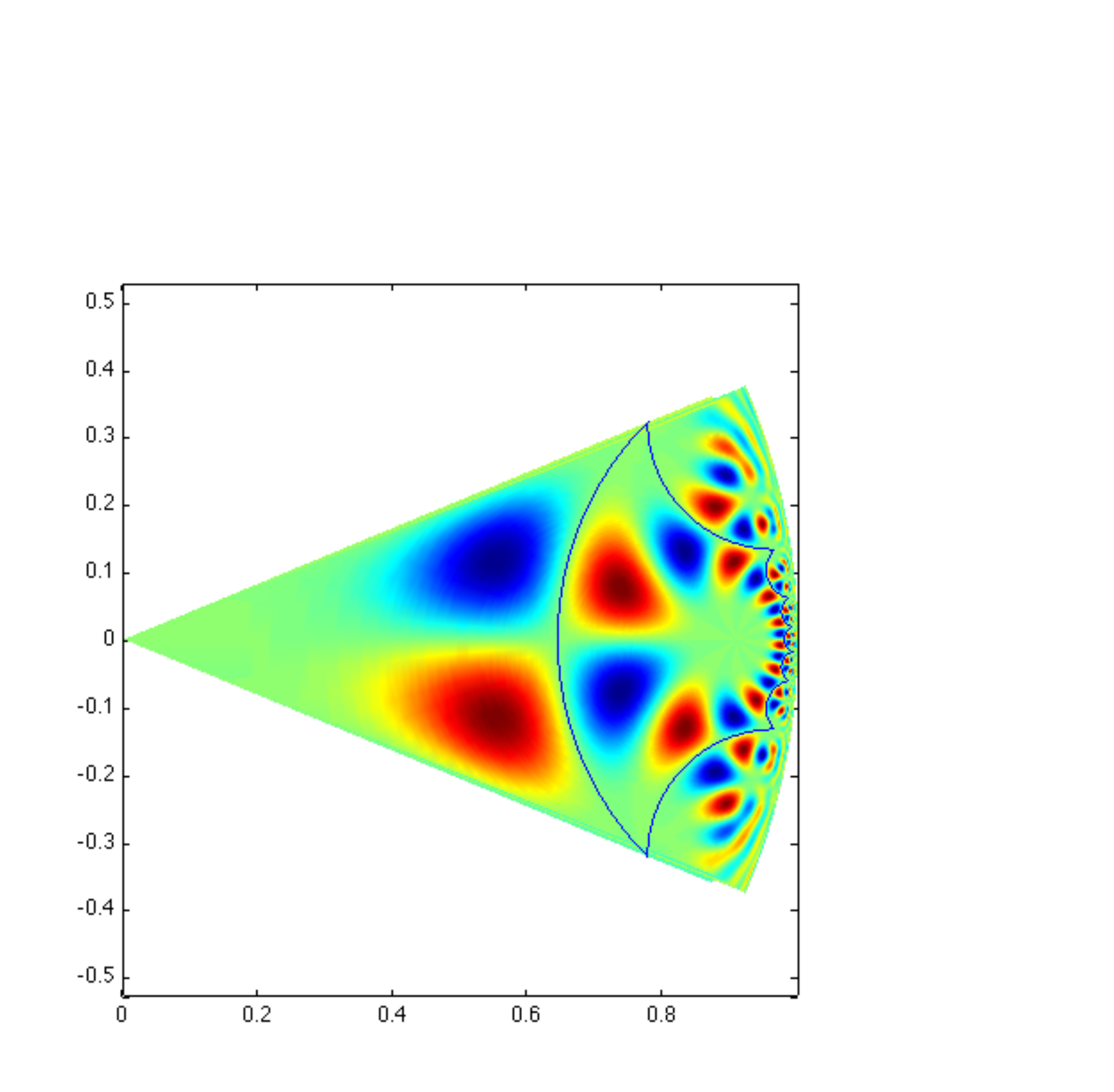}
}
\caption{\bf This is a zoom on the first ``octant'' in figure \ref{fig:image2_octogone}. It is at a higher spatial resolution than this figure for the second octagon, the one to the right of the ``main'' one and shows better the relationship between the intensity patterns within the two octagons.}
\label{fig:image_g0octogone}
\end{figure}

In order to illustrate what an eigenfunction for the regular octagonal group does look like, we have computed one such eigenfunction following the method exposed in \cite{balazs-voros:86}. The result is shown in Figure \ref{fig:image2_octogone}. Note the pattern which consists of pairs of blue and red spots  uniformly distributed around the central octagon (which is materialized by a dark line as well as the image under the generator $g_0$ of this octagon). This pattern is reproduced at infinity toward the boundary of the disc (which, in hyperbolic geometry, is at infinity) by acting with the elements of  $\Gamma_8$. In this figure the resolution becomes rapidly bad when approaching the boundary, but in Figure \ref{fig:image_g0octogone} we show a magnification of the sector in which the transformed octagon under $g_0$ lies. In this figure we can nicely see how the pattern inside the central octagon has been transformed under $g_0$. If one is interested in the interpretation of these images in terms of structure tensors rather than in terms of points in the Poincaré disk, one can use the ``$D$ to Tensor dictionary'' defined by equations \eqref{eq:DtoTensor}. As an example, looking at figure \ref{fig:image_g0octogone}, we see that the centers $z$ and $z'$ of the red and blue blobs in the ``main octagon'' are symmetric with respect to the horizontal axis and such that $z=0.55+0.1\,i$ and $z'=0.55- 0.1\,i$. This corresponds to the two structure tensors
\[
 \T=\left[
\begin{array}{cc}
 3.51 & 0.29\\
0.29 & 0.31
\end{array}
\right] \quad 
\T'=\left[
\begin{array}{cc}
 3.51 & -0.29\\
-0.29 & 0.31
\end{array}
\right],
\]
whose distance is equal to 0.81.

We should now take into account the symmetry group of the octagon, isomorphic to the dihedral group $D_8$ which contains 16 elements generated by the rotation $r\overset{\rm def}{=} r_{\frac{\pi}{4}}$ and by the reflection $\kappa$ through an axis of symmetry of the octagon. These transformations are all elements of ${\rm U}(1,1)$. The fundamental domain of $D_8$ in the octagon is $1/16$th piece of the cake. 
It follows from the calculations of \cite{balazs-voros:86} that the eigenvalues of $\Delta$ in this fundamental domain (with suitable boundary conditions) are simple, therefore the eigenvalues in the octogon with suitable periodic boundary conditions are either simple or double depending on the way in which the rotation $r$ acts on these eigenvectors. From the bifurcation point of view, this means that we may look for solutions in $D$ which are invariant under the action of $\Gamma_8$ and which transform like these eigenvectors under the action of $D_8$, henceforth reducing the problem to a simple or double eigenvalue problem with $D_8$ symmetry. 

The theory of $D_n$ symmetry breaking bifurcations ($n$ an integer) is well established, see \cite{golubitsky-stewart-etal:88}. We list below the generic situations which can occur according to the type of action of rotations and reflections in $D_8$ on the eigenvectors at a critical parameter value. We show in table \ref{table:generic} the generic bifurcations of $\Gamma_8$-periodic patterns. We note $\zeta$ an eigenvector of the Laplace-Beltrami operator $\Delta$ at a critical parameter value. Note that the octagon has two different types of symmetry axes: those joining opposite vertices and those joining the middle of opposite edges. The first case corresponds to points which are fixed under the reflection $\kappa$ (or a conjugate of $\kappa$ in $D_8$). The second case corresponds to points which are fixed under the reflection $\kappa'=r_{\frac{\pi}{8}} \kappa r_{\frac{\pi}{8}}$ (or a conjugate of $\kappa'$ in $D_8$).

Note that the periodic pattern illustrated in Figure \ref{fig:image2_octogone} corresponds to what a bifurcated state would look like in the case of the second line of table \ref{table:generic}.

We are however unable at this stage to tell without further and quite involved computations, which type of symmetry breaking will occur as the parameter $\mu$ crosses the stability threshold.

\section*{Discussion}
Our investigations are somewhat related to some of the issues raised  by Ermentrout \cite{ermentrout:98}. They are also related to the work of Bressloff, Cowan, Golubitsky, Thomas and Wiener \cite{bressloff-cowan-etal:01,bressloff-cowan-etal:02} on a model where either the connectivity kernel $w$ does not depend at all on the image features or is only sensitive to the (local) direction of the lines in it. This has led to beautiful results on the "spontaneous" occurence of hallucinatory patterns under the influence of psychotropic drugs. In further studies, Bressloff and Cowan have attempted to extend the theory to models taking into account not only the directional feature but also the spatial frequency in the images \cite{bressloff-cowan:02,bressloff-cowan:02b,bressloff-cowan:03}. Based on the experimental observation that hypercolumns seem to be organized around "pinwheels" in the visual cortex (points at which neurons are sensitive to any direction), they derived a model where direction and frequency define a point on the unit sphere $S^2$ and the connectivity kernel is invariant under the group ${\rm SO}(3)$ of rotations of the sphere. 

Our approach differs in that we model edges {\em and} textures simultaneously at a given scale through the structure tensor. The underlying feature space and its transformations are more complicated than the sphere $S^2$ and its rotation group ${\rm SO}(3)$. We showed that they can be represented by the Poincaré disk and its group of hyperbolic isometries. This naturally leads to a model of visual edges and textures where the equations are invariant by isometries in the (hyperbolic) space of structure tensors. Spatial scale can probably be included as well, this is the subject of future work.

There are also connections between our work and some previous work by Ben-Shahar, Zucker and colleagues \cite{ben-shahar-huggins-etal:03} who discuss the representation and processing in V1 of a larger set of visual features including edges, textures, shading, stereo. They do not deal at all with the problems of group invariance and of bifurcations of neural states, most likely because their underlying mathematical machinery, relaxation labelling \cite{faugeras-berthod:80b,hummel-zucker:83}, cannot easily address these questions. Ben-Shahar and Zucker pursue these ideas of ``good continuation'' of the texture flow from a more engineering viewpoint in \cite{ben-shahar-zucker:03}
and in \cite{ben-shahar-zucker:04} from the viewpoint of differential geometry as beautifully described in the book by Petitot \cite{petitot:09} and in some of his earlier papers \cite{petitot:03b}. It is clear that these complementary approaches should be brought together at some point and unified but this is the subject of future work.


The previous analyses and results use the assumption that the average voltage $V(\T,\gt)$ is invariant with respect to the action of the subgroup $N$ of ${\rm SU}(1,1)$. Thanks to this hypothesis we were able to reduce the dimension of the neural mass equation \eqref{eq:neuralmass} from 2 to 1 and to use classical Fourier analysis to describe the process of pattern formation and of bifurcation of the solutions.

One may argue that the action of the subgroup $\tilde{N}$ on the set of structure tensors does not have a natural interpretation, unlike that of $\tilde{K}$ and $\tilde{A}$ and, for that matter, that of $\tilde{\Gamma}_{n,t}$. On the other hand the subgroup $N$ features a very simple set of invariant functions, the H-planforms that can be used to represent the solutions of \eqref{eq:neuralmass} that are invariant with respect to its action. As far as we know similar functions are {\em not} known for the groups $\Gamma_{n,t}$ whose action on the set of structure tensors  {\em does} have a nice interpretation. This implies that the putative invariance of the average voltage $V(\T,\gt)$ with respect to this action would be most interesting to test through an analysis of the bifurcations of the solutions of \eqref{eq:neuralmassD} in the line of what we did for the group $N$ but is currently hampered by the lack of good functions for representing these solutions.

Another remark is that the "energy density" of these solutions tends exponentially fast to $\infty$ as $t$ tends to $-\infty$, due to the $e^{-2t}$ term in the expression of the hyperbolic surface element in horocyclic coordinates, see equation \eqref{eq:surface}. Such solutions may therefore not be physically admissible. This objection drops out for the H-planforms of the form $e^{(1+i\lambda)\langle z,b\rangle}$ with $\lambda\in\R$, as noted previously. Unfortunately one cannot carry out a simple bifurcation analysis for these H-planforms.

On the other hand we have seen above that such H-planforms can be associated, in a non trivial way, to periodic patterns with respect to the action of a discrete subgroup of ${\rm U}(1,1)$. This problem needs further investigation. The preliminary discussion about the octagonal group could a priori be transposed to many other kinds of hyperbolic patterns, and we do not know which one would be preferred, if any.

These examples are a few among many of an analysis that would have important implications in terms of the actual neural representation of the structure tensor (and at bottom of the image intensity derivatives). For example, given a subgroup $\Gamma$ of ${\rm SU}(1,1)$, assume that the mathematical analysis of the bifurcations of the solutions of equation \eqref{eq:neuralmassD} that are invariant with respect to the action of $\Gamma$ predicts the formation of certain patterns having the kind of symmetries represented by $\Gamma$. If such patterns can indeed be observed by actual measurements, e.g., optical imaging \cite{grinvald-hildesheim:04}, then this would be a strong indication that the neural ``hardware'' is built in such a way that its state is insensitive to the action of $\Gamma$. For example, in equation \eqref{eq:neuralmassD}, the state is the average membrane potential $V(z,\tau)$. The observation of the above pattern formation would come in support of the hypothesis that $V(\gamma \cdot z,\tau)=V(z, \gt)$ for all elements $\gamma$ of the group $\Gamma$, for all structure tensors $z$ and for all time instants $\gt$. In other words, bifurcation theory and pattern formation can be considered as theoretical probes of various hypotheses about the neural organization of the brain, allowing to make precise predictions about the kinds of patterns that should be observed in the activity of real brains, and opening the door to the design of experiments to test these hypotheses. Specific examples of such groups are the groups $\Gamma_{n,T}$ we gave a few examples of and the octagonal group $\Gamma_8$ discussed previously.

The restriction to the hyperbolic plane instead of the three-dimensional space of structure tensors looks like an oversimplification, which should be only considered a useful first step. Our plan is to extend this analysis to the full tensor space, making use if necessary (and this will certainly be the case) of numerical simulations in order to get a better idea of the phenomenology.

As mentioned in the Methods Section, it is natural to consider a spatial extension of our analysis that would analyze a spatial distribution of the kind of structure tensor hypercolumns that we have described in this paper, see equation \eqref{eq:neuralfield}. This would lead in particular to an analysis of ``hyperbolic hallucinatory patterns'' that could be compared against those described in the work of Bressloff, Cowan, Golubitsky and collaborators \cite{bressloff-cowan-etal:01,bressloff-cowan-etal:02}. This requires first to better understand our a-spatial model and is the subject of some of our future investigations.

One may also speculate what such an array of structure tensors would offer compared to an array of orientations. Even if this has not yet been worked out to our knowledge in the context of neural fields, it is likely that an array of orientations can support the perception of extended contours in an otherwise ``flat'' image, like a cartoon \cite{field-hayes-etal:93,field-hayes:04}. This can be achieved by such connectivity functions as those that enforce the Gestalt law of good continuation. As mentioned above some of these ideas can be found in the work of Steve Zucker and his associates. An array of structure tensors would add to this the possibility of perceiving extended texture edges such as those encountered in natural images where sharp variations in the texture are likely to indicate boundaries between objects. This is certainly a very important area of investigation from the psychophysical, neurophysiological and mathematical perpectives.

A final remark is that all this analysis assumes a perfectly invariant problem under the group of isometries in the space of structure tensors, a situation which is of course very unlikely, but which has the great advantage to allow for computations and to highlight fundamental properties and features of the problem at hand. A next step would be to look at the "imperfect" case in which symmetries are not perfectly satisfied, but this, even in the simplified context of the Poincar\'e disc, may be a formidable challenge. 
 \\
{\bf Acknowledgement}\\
This work was supported by the ERC advanced grant NerVi \#227747.
\section*{Supporting Information}
\subsection*{Supplementary text S1}
We describe the relation between ${\rm SDP}(2)$ and ${\rm SSDP}(2)$.

By identifying  ${\rm SDP}(2)$ with the quotient ${\rm GL}(2,\R)/{\rm O}(2)$, we see that it is also a homogeneous space of the Lie group ${\rm GL}(2,\R)$ of $2\times 2$ invertible matrices with real coefficients.
 It is useful to consider the symmetric space of special symmetric positive matrices ${\rm SSDP}(2) = {\rm SDP}(2) \cap {\rm SL}(2,\R)= \{A \in {\rm SDP}(2), {\rm det} A = 1\}$.
This submanifold can also be identified with the quotient ${\rm SL}(2,\R)/{\rm SO}(2)$, which is itself isomorphic to the hyperbolic space $H_2$. Here ${\rm SL}(2,\R)$ denotes the special linear group of all determinant one matrices in ${\rm GL}(2, \R)$. Therefore ${\rm SSDP}(2)$ is a totally geodesic submanifold of ${\rm SDP}(2)$ \cite{maass:71}. Now
since ${\rm SDP}(2)={\rm SSDP}(2) \times \R^+$, it can be seen as a foliated manifold whose codimension-one leaves are isomorphic to the hyperbolic surface $H^2$.

\subsection*{Supplementary text S2}
The Poincaré half-plane model, noted $\mathcal H$, is obtained from the Poincaré disk model by the mapping $f$ such that
\[
 u=f(z)=-i \frac{z+1}{z-1}
\]
which is an isometry from  $D$ to the upper half-plane $\mathcal{H}:~\{{\rm Im}(z) > 0\}$. The distance between two points $u$, $u'$ in $\mathcal{H}$ is then easily obtained from the distance in $D$ by setting $z=f^{-1}(u)$ and $z'=f^{-1}(u')$ in the expression (\ref{d2}). This gives
\begin{equation}\label{eq:d3}
d_3(u,u') = d_2(f^{-1}(u),f^{-1}(u'))= \arctanh \frac{|u'-u|}{|u'-\overline u|}
\end{equation}
Geodesics in $\mathcal{H}$ are lines or circles orthogonal to the real axis.
The surface element in $H^2$ is 
$$ds^2=u_2^{-2}(du_1^2+du_2^2),$$
if $u=u_1+iu_2$.
\subsection*{Supplementary text S3}
We describe a spherical model for the set SDP(2) of structure tensors.
The constraint that the determinant of the structure tensor should be equal to 1 is unnatural since in a given image the values of the structure tensors determinants are likely to vary over a wide range. We saw that the set of structure tensors, SDP(2), was a foliated manifold whose co-dimension 1 leaves are isomorphic to $H^2$. We can also represent ${\rm SDP}(2)$ as the open unit ball of $\R^3$.

Let
\[
 \T=\left[ \begin{array}{cc}
 a & c \\
c & b
\end{array}
\right]
\]
be an element of ${\rm SDP}(2)$ of determinant equal to $ab-c^2=d^2 \geq 0$. The change of variables
\[
 x_0=\frac{a+b}{2}\quad x_1=\frac{a-b}{2}\quad x_2=c,
\]
indicates that all tensors of determinant equal to $d^2$ belong to the sheet of the hyperboloid of equation
\[
 x_0^2-x_1^2-x_2^2=d^2
\]
corresponding to positive values of $x_0$. If we perform the stereoscopic projection of this sheet with respect to the point of coordinates $(0,0,-d)$ in the plane of equation $x_0=0$ one obtains the open disc of radius $d \geq 0$.

Consider now the subset of tensors with determinant less than or equal to 1 ($0 \leq d \leq 1$). For each $d$ we have a one to one correspondence between the tensors of determinant equal to $d^2$ and the points of the  open disk in the plane of equation $x_0=0$ centered at the origin and of radius equal to $d$, hence with the same open disk centered on the $x_0$-axis but in the plane of equation $x_0=1-d$. This establishes a one to one correspondence between the tensors of determinant between 0 and 1 (including these values) and the northern half open unit ball of center the origin.

Consider next the subset of tensors with determinant greater than or equal to 1 ($d \geq 1$). The inverse of each such tensor has a determinant equal to $0 < 1/d \leq 1$. We have therefore a one to one correspondence between the set of tensors of determinant $d^2$ and the points of the open disk of radius $1/d$ in the plane of equation $x_0=1/d-1$ centered on the $x_0$-axis. This establishes a one to one correspondence between the tensors of determinant greater than or equal to 1  and the southern half open unit ball of center the origin.

\begin{figure}[htbp]
 \centerline{\includegraphics[width=0.5\textwidth]{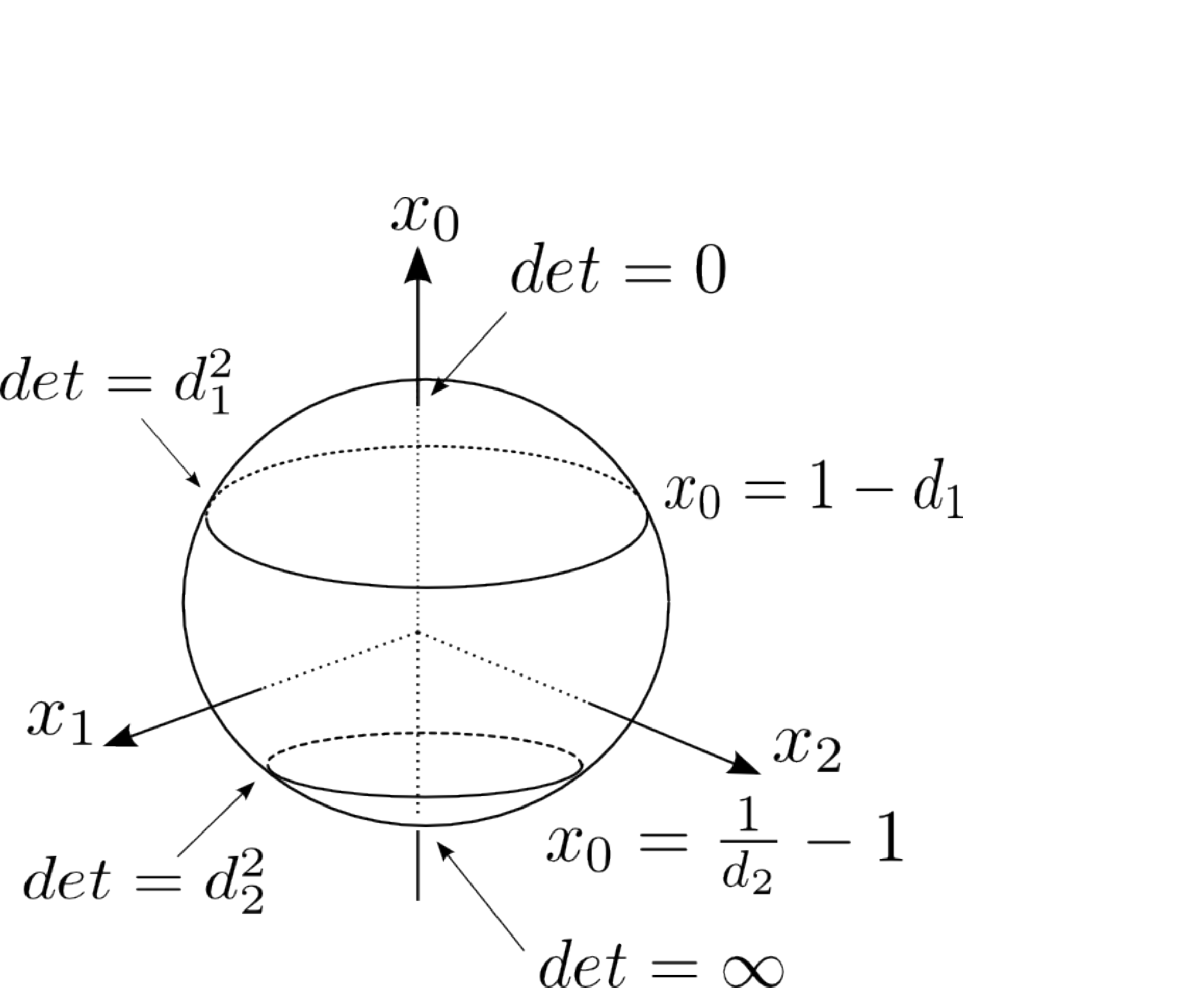}}
\caption{{\bf The unit ball of $\R^3$ is a model of SDP(2), see text in the supplementary material text S3.}}
 \label{fig:sphere}
\end{figure}

Combining these two representations we obtain a one to one correspondence between the set SDP(2) of structure tensor and the open unit ball centered at the origin, see figure \ref{fig:sphere}.

This representation has the following nice property. If $\T$ is an element of SSDP(2), $d\,\T$, $d>0$ is an element of SDP(2) with determinant $d^2$. Let $m$ be the point representing $\T$ and $P$ that representing $d\,\T$. An easy verification shows that the projection of $p$ of $P$ in the $x_0$-plane is obtained by applying the homotethy of center the origin and of ratio $d$ to the point $m$.

The Riemannian structure of SSDP(2) is transported to the open unit ball as follows. Consider two structure tensors $\T_1$ and $\T_2$ with determinants $d_1^2$ and $d_2^2$. Define $\overline{\T}_i=\frac{1}{d_i} \T_i$, $i=1,2$ that are in SSDP(2). The geodesic $\mathcal{G}(t)$ between $\T_1$ and $\T_2$ can be parameterized by \cite{moakher:05}
\[
 \mathcal{G}: [0,1] \to {\rm SDP}(2) \quad \text{such that} \quad \mathcal{G}(t)=\T_1^{1/2} e^{t\log \left(\T_1^{-1/2}\, \T_2 \,\T_1^{-1/2}\right)} \T_1^{1/2}
\]
A simple algebraic manipulation shows that
\[
 \mathcal{G}(t)=d_1^{1-t} d_2^t \,\overline{\T}_1^{1/2} e^{t\log \left(\overline{\T}_1^{-1/2}\, \overline{\T}_2 \,\overline{\T}_1^{-1/2}\right)} \overline{\T}_1^{1/2}=d_1^{1-t} d_2^t \, \overline{\mathcal{G}}(t),
\]
where $\overline{\mathcal{G}}(t)$ is the geodesic in SSDP(2) between $\overline{\T}_1$ and $\overline{\T}_2$. In the sphere model the corresponding geodesic is obtained very simply as follows. Let $m_1$ and $m_2$ be the two points of the open unit disk
centered at the origin in the plane of equation $x_0=0$ (this is the representation of SSDP(2)). The geodesic between $m_1$ and $m_2$ is the circular arc  going through $m_1$ and $m_2$ orthogonal to the unit circle. Let $m_t$ be the point of this geodesic representing $\overline{\mathcal{G}}(t)$. When $t$ varies from 0 to 1, the point $m_t$ traces the geodesic arc between $m_1$ and $m_2$. According to a previous remark, the projection in the $(x_1,x_2)$ plane of the point $P_t$ representing the tensor $\mathcal{G}(t)$ is obtained by applying the homotethy of center the origin and ratio $d(t)=d_1^{1-t} d_2^t$ to $m_t$ and its $x_0$-coordinate is $1-d(t)$ if $d(t) \leq 1$ and $1/d(t)-1$ if $d(t) \geq 1$.
\subsection*{Supplementary text S4}
We prove the following proposition that is stated without proof in the Section Methods:
\begin{proposition*}
 $\Gamma_{2,T}$ is a Fuchsian group for all $T \neq 0$. $\Gamma_{4,T}$ (respectively $\Gamma_{6,T}$) is a Fuchsian group if $\cosh T \geq \sqrt{2}$ (respectively
if $\cosh T \geq 2$).
\end{proposition*}
\begin{proof}
According to  \cite[chapter 2]{katok:92}, in order to prove that $\Gamma_{n,t}$ is Fuchsian it is sufficient to prove that it is a discrete subgroup of ${\rm SU}(1,1)$. Since $\Gamma_{n,t}$ is the free product of the two cyclic groups $K_n$ and $A_T$. Theorem 1 in \cite{rosenberger:72} gives a necessary and sufficient condition for such a subgroup of ${\rm SU}(1,1,)$ to be discrete. We define
$\lambda_p=2\cos \frac{\pi}{p}$, $p \geq 2$. Rosenberger's first theorem states that a sufficient condition for a free group product $G$ of two cyclic subgroups of ${\rm SU}(1,1)$ is that there exist two generators $U$ and $V$ such that
\begin{itemize}
 	\item ${\rm Tr}(U)=\lambda_p$ or ${\rm Tr}(U) \geq 2$, ${\rm Tr}(V)=\lambda_q$ or ${\rm Tr}(V) \geq 2$,
	\item $UV \neq \pm {\rm Id}$ when ${\rm Tr}(U)={\rm Tr}(V)=0$,
	\item ${\rm Tr}(UV^{-1}) \leq -2$.
\end{itemize}
Let $r_{2\pi/n}$ be the element of $K_{n}$ corresponding to the rotation of angle $\pi/n$, $n=2,4,6$. It is clear that $\Gamma_{n,t}$ is generated by
the pair $(r_{2\pi/n},a_T)$ and that ${\rm Tr}(r_{2\pi/n})=\lambda_n$ and ${\rm Tr}(a_T)=2 \cosh t$. On the other hand ${\rm Tr}(r_{2\pi/n}(a_T)^{-1})=\lambda_n \cosh t$ which does not allow us to conclude. 

Consider the case $n=2$ and note that $K_{2, t}$ is also generated by the pair $(U_2^{T},V_2^{T})=(r_{\pi},r_{\pi}^{-1} a_T)=(r_{\pi},r_{-\pi} a_T)$. It is easy to check that ${\rm Tr}(U_2^{T})=\lambda_2=0$, ${\rm Tr}(V_2^{T})=\lambda_2 \cosh T=0$, $U_2^{T}V_2^{T}=a_T\neq {\rm Id}$ if $ T \neq 0$ and ${\rm Tr}(U_2^{T} (V_2^{T})^{-1})=-2 \cosh T \leq -2$ for all $T$s.

Consider the case $n=4$ and note that $K_{4,T}$ is generated by the pair $(U_4^{T},V_4^{T})=(r_{\pi/2},r_{\pi/2}^{-2} a_T)=(r_{\pi/2},r_{-\pi/2} a_T)$. It is straightforward to check that  ${\rm Tr}(U_4^{T})=\lambda_4$, ${\rm Tr}(V_4^{T})=\lambda_2 \cosh T=0$ and that ${\rm Tr}(U_4^{T} (V_4^{T})^{-1})= 2 \cos \frac{3 \pi}{4} \cosh T=-\sqrt{2} \cosh T$. Thus $K_{4,T}$ is Fuchsian if $\cosh T \geq \sqrt{2}$.

Consider finally the case $n=6$ and note that $K_{6,t}$ is generated by the pair $(U_{6}^{T},V_{6}^{T})=(r_{\pi/6},r_{\pi/6}^{-3} a_T)=(r_{\pi/6},r_{-\pi/2} a_T)$. It is straightforward to check that  ${\rm Tr}(U_{6}^{T})=\lambda_6$, ${\rm Tr}(V_{6}^{T})=\lambda_2 \cosh T=0$ and that ${\rm Tr}(U_{6}^{T} (V_{6}^{T})^{-1})= 2 \cos \frac{2 \pi}{3} \cosh T=- \cosh T$. Thus $K_{6,T}$ is Fuchsian if $\cosh T \geq 2$.
\end{proof}

\newpage
\section*{Tables}
\begin{table}[!ht]
 \caption{{\bf A glossary of mathematical notations}}
\begin{tabular}{|l|l|}
\hline
 ${\rm SDP}(2,\R)$  & The set of two-dimensional symmetric definite positive real matrixes.\\\hline
${\rm SSDP}(2,\R)$ & The subset of ${\rm SDP}(2,\R)$ whose elements have a determinant equal to 1.\\\hline
${\rm U}(1,1)$ & The indefinite unitary group of two-dimensional complex matrixes that\\ \hline
		& leave invariant the sesquilinear form $|z_1|^2-|z_2|^2$.\\\hline
${\rm SU}(1,1)$ & The subgroup of ${\rm U}(1,1)$ whose elements have a determinant equal to 1.\\\hline
${\rm GL}(2,\R)$ & The group of two-dimensional invertible real matrixes.\\\hline
${\rm SL}(2,\R)$ & The special linear group of two-dimensional real matrixes with determinant equal to 1.\\\hline
${\rm E}(2,\R)$ & The group of Euclidean transformations of $\R^2$.\\\hline
${\rm O}(2)$ & The group of two-dimensional real orthogonal matrixes.\\\hline
${\rm SO}(2)$ & The special orthogonal group of the real orthogonal matrixes with determinant equal to 1.\\\hline
$D_4$ & The symmetry group of a square.\\\hline
$D_6$ & The symmetry group of the hexagon.\\\hline
$D_8$ & The symmetry group of the octagon.\\\hline
$D$ & The open disk of radius 1.\\\hline
$\partial D$ & The boundary of $D$, the unit circle.\\\hline
$\mathcal H$ & The hyperbolic space.\\\hline
\end{tabular}
\label{table:glossary}
\end{table}

\begin{table}[!ht]
 \caption{{\bf Generic bifurcations of $\Gamma_8$-periodic patterns.} Each case in the table  corresponds to an irreducible representation of the group $D_8$.}
\begin{tabular}{|l|l|}
\hline
$D_8$ acts trivially on $\zeta$ & simple eigenvalue, transcritical branch of states \\
 & with full $D_8$ symmetry\\\hline
$r\cdot\zeta=\zeta$, $\kappa\cdot\zeta=-\zeta$ &  simple eigenvalue, pitchfork branch of rotationally\\
and $\kappa'\cdot\zeta=-\zeta$ & invariant states with broken $\kappa$, $\kappa'$ symmetry
 \\\hline
$r\cdot\zeta=-\zeta$ and either & simple eigenvalue, pitchfork branch of states  \\
$\kappa\cdot\zeta=-\zeta$ or $\kappa'\cdot\zeta=-\zeta$ &  with partially broken rotational symmetry \\
 & (since $r^2\cdot\zeta=\zeta$ the state keeps a 4-folds symmetry)\\\hline
$r\cdot\zeta=\zeta'$ where $\zeta'$  & several subcases can occur, for example if $r\zeta'=-\zeta$ the problem is similar \\
 is not colinear to $\zeta$ & to one with $D_4$ symmetry breaking. \\
 & The critical eigenvalue is double, rotational symmetry is broken and \\
  & there are generically two pitchfork branches of bifurcated solutions: \\
& those which keep the symmetry under reflection $\kappa$ and \\
& those which keep the symmetry under $\kappa'$\\\hline
\end{tabular}
\label{table:generic}
\end{table}

\newpage

\end{document}